\documentclass[12pt]{article}
\usepackage[paperwidth=21cm,paperheight=26cm, textwidth=16cm,textheight=20.6cm]{geometry}
\usepackage{amsmath,amsthm,amssymb,xspace,xcolor,hyperref,bm}\hypersetup{colorlinks=true, linkcolor=black,citecolor=black,urlcolor=black}\usepackage{graphicx}
\numberwithin{equation}{section}

\theoremstyle{plain}
\newtheorem{theorem}{Theorem}

\newtheorem{claim}{Claim}[section]
\newtheorem{proposition}{Proposition}
\newtheorem{lemma}{Lemma}
\newtheorem{corollary}{Corollary}
\newtheorem{definition}{Definition}

\newtheorem{problem}{Problem}

\newcommand{\APC}{\ensuremath{{\sf APC_1}}\xspace}

\newcommand{\SB}{\ensuremath{{\sf S^1_2}}\xspace}

\newcommand{\Ppoly}{\ensuremath{{\sf P/poly}}\xspace}
\newcommand{\NC}{\ensuremath{{\sf NC}^1}\xspace}
\newcommand{\AC}{\ensuremath{{\sf AC}^0}\xspace}

\newcommand{\NP}{\ensuremath{{\sf NP}}\xspace}

\newcommand{\Circuit}{\ensuremath{{\sf Circuit}}\xspace}

\newcommand{\SAT}{\ensuremath{{\sf SAT}}\xspace}

\newcommand{\MCSP}{\ensuremath{{\sf MCSP}}\xspace}

\begin{document}

\title{{\bf Localizability of the approximation method}}
\author{J\'an Pich \\ \small University of Oxford}
\date{\small December 2022}
\maketitle

\begin{abstract}
We use the approximation method of Razborov to analyze 
 the locality barrier which arose from the investigation of the hardness magnification approach to complexity lower bounds. Adapting a limitation of the approximation method obtained by Razborov, we show that in many cases 
 it is not possible to combine the 
 approximation method with typical (localizable) hardness magnification theorems to derive strong circuit lower bounds. In particular, one cannot use the 
 approximation method to derive an extremely strong constant-depth circuit lower bound and then magnify it to an \NC lower bound for an explicit function.

To prove this we show that
 lower bounds obtained by the 
 approximation method are in many cases localizable in the sense that they imply lower bounds for circuits which are allowed to use
 arbitrarily powerful oracles with small fan-in. 
\end{abstract}


\section{Introduction}

\noindent {\bf Approximation method.} In the 1980s Razborov \cite{Rmclb1,Rmclb2} initiated an approach to
 proving lower bounds on the size of Boolean circuits known as the {\em approximation method}. The idea of the  
 method was to approximate each small circuit by an `approximating' circuit, by replacing each connective of the original circuit by an `approximating' connective, and then show that the approximating circuits are too weak to compute a given Boolean function $f$. Consequently, each small 
 circuit had to make an error in computing $f$ as well. This strategy 
was successfully implemented in lower bounds for restricted classes of circuits such as monotone circuits \cite{Rmclb1,Rmclb2} or circuits of constant depth \cite{Rac0,Sac0}.

The formal framework of the approximation method was suitable also for a systematic analysis of potential ways of proving circuit lower bounds. 
In \cite{Ram1} Razborov gave a remarkable example of such meta-analysis of circuit complexity by showing 
 that the approximation method cannot prove better than $O(nn_0)$-size lower bounds for general circuits, for 
 any Boolean function $f$, where $n_0$ is the number of essential variables of $f$. The essential variables are the variables on which the function depends. Moreover, if we consider a particular form of the approximation method 
 which was used to derive lower bounds for monotone and constant-depth circuits, the best lower bound we can hope for is $24(n_0+1)$. Despite these limitations, Razborov \cite{Ram1} showed that the approximation method is in certain sense complete: 
 If a Boolean function $f$ is not computable by any circuit of size $s^3$, one can use the approximation method to prove an $\Omega(s)$-size circuit lower bound for $f$. 
 This requires introducing many inessential variables and proving the lower bound for the new function with extra variables. The complete version of the approximation method 
 became known as the fusion method \cite{Wfus}.
\bigskip

\noindent {\bf Natural proofs.} 
 Unfortunately, the promise of the fusion method has not been fulfilled as strong complexity lower bounds for explicit Boolean functions remain elusive. A significant part of the reason for this situation can be attributed to 
 the natural proofs barrier of Razborov and Rudich discovered in the
1990s \cite{RR}. Intuitively, natural proofs say that the existing circuit lower bounds do not prove just that a particular function requires big circuits but they yield even efficient algorithms rejecting all easy functions while accepting many hard functions. 
 The existence of such an algorithm for general Boolean circuits (formally, a \Ppoly-natural property against \Ppoly) 
 would, however, break cryptographic pseudorandom generators. 
\bigskip

\noindent {\bf Hardness magnification.} A generic approach to lower bounds which seems to avoid 
 the natural proofs barrier was investigated in more recent years in the area of {\em hardness magnification}. Hardness magnification is an approach to strong complexity lower bounds by reducing them to lower bounds against weak computational models. While some forms of magnification can be traced back to early 2000s, the term was coined by Oliveira and Santhanam \cite{HM} in 2018 to refer to its newer instantiations, see \cite{CHOPRS} for a more comprehensive exposition. In particular, some subsequent papers established so called HM frontiers, which refer to theorems including three statements of the following kind.

\vspace{0.3cm}

\setlength\fboxrule{0.6pt}
\hspace{-10pt}\fbox{\parbox{430pt}{
\vspace{6pt}
\quad {\bf HM frontier for} $\mathsf{AC}^0$-$\mathsf{XOR}$\textbf{:}
\medskip

\quad 1. If $\MCSP[n^c,2n^{c}] \notin \mathsf{AC}^0$-$\mathsf{XOR}[N^{1.01}]$ for some $c$, then $\mathsf{EXP} \nsubseteq \mathsf{NC}^1$ \cite[\S 3.2.2]{CHOPRS}.

\quad 2. $\mathsf{MAJ} \notin \mathsf{AC}^0$-$\mathsf{XOR}[2^{n^{o(1)}}]$, cf. \cite{Rac0,Sac0}.

\quad 3. $\MCSP[n^c,2n^{c}] \notin \mathsf{AC}^0$, 
 if $c$ is sufficiently large \cite[Theorem 52]{CHOPRS}.
\vspace{3pt}}}

\vspace{0.3cm}

\noindent Here, $\MCSP[s,t]$ is the promise problem of determining if an $N:=2^n$-bit input has circuit complexity at most $s$ versus at least $t$. $\mathsf{AC}^0$-$\mathsf{XOR}[s]$ denotes 
 $s$-size constant-depth circuits with {\sf XOR} gates at the bottom layer, where the {\sf XOR} gates compute 
parity functions with arbitrary fan-in. 

Items 2 and 3 thus indicate that proving the lower bound required in Item 1 could be within reach as both the corresponding circuit class and computational problem are prone to a nontrivial analysis. HM frontiers can be established for many computational models instead of \AC-{\sf XOR} and for various conclusions in Item 1, e.g. $\NP\not\subseteq\NC$ \cite{CHOPRS}. 

The reason why 
 magnification theorems seem to overcome the natural proofs barrier is that Item 1 relies on specific properties of the function in question (in our case, \MCSP), it is not clear how to generalize it to a substantial fraction of functions. In fact, this intuition was formally supported in \cite{CHOPRS} by showing that hardness magnification is in certain cases inherently nonnaturalizable in the sense that the conclusion in Item 1 is the non-existence of natural proofs against \Ppoly. 
\bigskip

\noindent {\bf Locality barrier.} Unsurprisingly, hardness magnification suffers from its own barrier, the {\em locality barrier}.\footnote{Lower bounds from Item 3 of HM frontiers are affected also by the black-box natural proofs barrier \cite{BBnatpr}.} As it turns out, Items 1 in HM frontiers are always obtained by constructing an oracle upper bound. 
For example, in our case, Item 1 follows from showing that $\MCSP[n^c,2n^{c}]$ can be computed by an $\AC$-{\sf XOR}$[O(N)]$-circuit using oracles with fan-in $N^{\epsilon}$, for some $\epsilon<1$.
 On the other hand, Items 2 and 3 can be generalized so that they work against circuits with such oracles. We refer to oracles with small fan-in as {\em local} oracles and to magnification theorems obtained by constructing oracle upper bounds as {\em localizable} magnifications.

The fact that many known circuit lower bounds can be extended to models allowing arbitrary local oracles is interesting independently of hardness magnification. Proving the non-existence of subexponential-size learning algorithms for \Ppoly would imply the non-existence of \Ppoly-natural properties against \Ppoly \cite{CIKK}, but it is not hard to see that natural properties against \Ppoly are computable by a single local oracle applied on a prefix of the input. Overcoming the locality barrier is thus essential for proving strong complexity lower bounds of our interest.

There are examples of non-localizable lower bounds but they suffer from drawbacks such as that they do not achieve an HM-frontier or they yield at best uniform lower bounds or `non-explicit' lower bounds of the form {\sf QP} $\not\subseteq\Ppoly$,
 where {\sf QP} stands for quasipolynomial time. We refer to \cite[\S 5]{Plc} for a more detailed discussion of these exceptions including non-localizable magnification theorems.
\bigskip

The motivating question of the present paper is to understand the extent of localizability of circuit lower bounds. Do `concrete' sufficiently strong complexity lower bounds always localize? We address the question by using the approximation method as a convenient substitute for the informal notion of a concrete lower bound.

\subsection{Our contribution}

Unlike Razborov \cite{Ram1,Ram2}, who investigated general properties of the approximation method for unrestricted circuits, we focus primarily on analyzing the approximation method for weak circuit classes such as \AC. This is motivated by hardness magnification. Nevertheless, we start by observing a couple of properties of the approximation method for unrestricted circuits.

\subsubsection{Limitations: Localizability}

Lower bounds based on the approximation method are proved by showing lower bounds for a measure $\rho(f,{\cal M})$ which expresses a distance of the target function $f$ for which we want to prove a lower bound and a suitable approximation model ${\cal M}$. 
 In fact, $\rho(f,{\cal M})$ is a lower bound on the circuit complexity of $f$. The above-mentioned barrier result of Razborov \cite{Ram1} says that for each legitimate model ${\cal M}$ and each Boolean function $f$, $\rho(f,{\cal M})\le O(nn_0)$. This leaves open the possibility of proving a superpolynomial circuit lower bound for $f$ by proving a superlinear lower bound for $\rho(h,{\cal M})$, for another Boolean function $h$, and then magnifying it. 
\smallskip

\noindent {\bf Localizability of general circuit lower bounds.} We observe that such attacks are ruled out for localizable magnification theorems under additional assumptions such as that the number of oracles in the oracle circuit implying the magnification theorem is sublinear. In other words, we observe that superlinear lower bounds for general circuits based on the
 approximation method yield lower bounds for general circuits with a limited number of local oracles, cf. Theorem \ref{t:genloc}. 
\smallskip

\noindent {\bf Limitations of the approximation method for constant-depth circuits.} Next, we consider the question of deriving \NC lower bounds via subexponential-size constant-depth circuit lower bounds based on the approximation method. We do not rule out such possibility but we show that Razborov's barrier result can be adapted to the case of constant-depth circuits with nontrivial conclusions. Informally, we show that for each legitimate model $\cal M$ and each Boolean function $f$, $\rho_d(f,{\cal M})\le O(2^{8n/d})$,
 cf. Theorem \ref{t:damlim}. Here, $\rho_d(f,{\cal M})$ denotes a version of $\rho$ for circuits of depth $d$. In particular, $\rho_d(f,{\cal M})$ is a lower bound on the size of $d$-depth circuits computing $f$.
\medskip

\noindent {\bf Localizability of constant-depth circuit lower bounds.} Our main theorem shows that localizability is an inherent property of all sufficiently strong lower bounds for circuits of constant depth obtained via the
 approximation method. 

\begin{theorem}[Localizability of the approximation method for constant-depth circuits - Informal, cf. Theorem \ref{t:dloc}]\label{t:idloc} Let $\cal M$ be any legitimate model of constant-depth circuits and $f$ any Boolean function with $n$ inputs. Suppose that $\rho_{O(d^2)}(f,{\cal M})\ge s$. Then $f$ is not computable by depth $d$ circuits of size $s-poly(kdn(2^{m/d}+1))$
 using $k$ arbitrarily powerful oracles of arity $m$. The constants in the $poly(\cdot)$-notation are absolute.\footnote{We work with a `symmetric' definition of $\rho_d$, cf. Definitions \ref{d:distance} and \ref{d:ddistance}, but Theorem \ref{t:idloc} holds for `asymmetric' $\rho_d$ as well, if we adjust the parameters appropriately.}
\end{theorem}

\noindent Note that showing that $f$ is not computable by $poly(n2^m)$-size circuits of depth $d$ implies that $f$ is not computable by p-size circuits of depth $d/3$ 
 with oracles of arity $m$ because we can replace such oracles by DNFs of size $2^{O(m)}$. Theorem \ref{t:idloc} shows that even $poly(n2^{m/d})$-size lower bounds localize, if they are obtained by the approximation method. 

\subsubsection{Completeness: Extent of Fusion} 

In order to prove an $s(n)$-size circuit lower bound for a Boolean function $f$ with $n$ inputs using the approximation method, it is necessary to introduce inessential variables, i.e. to consider function $f'(x_1,\dots,x_N):=f(x_1,\dots,x_n)$, for $N\ge s(n)$, and prove $\rho(f',{\cal M})\ge s(n)$. Razborov \cite{Ram1} showed that this strategy can be adapted (using the asymmetric definition of $\rho$) 
 so that it is complete. Wigderson \cite{Wfus}, following Karchmer's \cite{Kfus} interpretation of Razborov's construction, coined the term {\em fusion method} for the resulting approach to circuit lower bounds. 
\smallskip

\noindent {\bf Feasibility of fusion.} The construction from \cite{Ram1} uses $N=O(s2^n)$. We observe that the fusion method can be adapted to the case of partial Boolean functions so that the corresponding  approximation models use $N=poly(s)$, cf. Theorem \ref{t:afcomp}. The adaptation relies on the notion of anticheckers of Lipton and Young \cite{LY}. The resulting fusion method inherits completeness from the standard fusion method. If anticheckers can be generated efficiently (which can be done in the case of \SAT under standard hardness assumptions), the corresponding approximation models are constructive, see \S\ref{s:gfusion}.
\smallskip

\noindent {\bf Fusion for constant-depth circuits.} Finally, we define a version of the fusion method for constant-depth circuits and show that it is complete for constant-depth circuit lower bounds, cf. Theorem \ref{t:dfcomp}. 

\subsubsection{Proof methods} 

Informally, the measure $\rho(f,{\cal M})$ is defined as the minimal number of `error sets' needed to cover all inputs on which $f$ differs from its `approximator'. To get a good lower bound, one wants to define a model $\cal M$ with small error sets but with approximators making a lot of errors. If $f$ is computable by an $s$-size circuit $W$, the approximator of $W$ and error sets corresponding to the gates of $W$ can be used to witness that $\rho(f,{\cal M})\le s$.

Razborov's proof of $\rho(f,{\cal M})\le O(nn_0)$ proceeds by constructing suitable distributions on error sets and approximators, which allow us to derive the desired conclusion by a case analysis: For each input $x$, either I. the probability that a random approximator fails to compute $f$ on $x$ is small or II. the probability that a random error set covers $x$ is high. Then, we can take a majority of random approximators and obtain an approximator coinciding with $f$ on all inputs from case I. The remaining inputs from case II are covered by a small number of error sets. The construction of the distribution of error sets and approximators is based on an exponentially-big circuit which plays the role of circuit $W$ from the previous paragraph.

Theorem \ref{t:idloc} is proved by the contrapositive. We assume that there is a small constant-depth oracle circuit computing $f$ and we want to bound $\rho_{O(d^2)}(f,{\cal M})$. If the constant-depth circuit did not contain any oracles, its approximator would yield a small set of error sets covering all inputs on which the approximator fails to compute $f$. The problem is to deal with local oracles. 

The first step of the proof is to adapt Razborov's barrier to $\rho_d$, i.e. to show $\rho_d(h,{\cal M})\le O(2^{8n/d})$ for every Boolean function $h$ with $n$ inputs. To make this work we need to use suitable constant-depth circuits as an alternative to the exponential-size circuits in Razborov's construction.

Having $\rho_d(h,{\cal M})\le O(2^{8n/d})$ and the assumption that there is a small oracle circuit $C$ for $f$, we want to obtain an upper bound on $\rho_{O(d^2)}(f,{\cal M})$. Intuitively, we want to replace each oracle in $C$ computing a function $h$ with $m$ inputs by $\rho_d(h,{\cal M})\le O(2^{8m/d})$. Unfortunately, we do not have a generic way to derive such upper bound just from an upper bound on $\rho_d(h,{\cal M})$. To make the proof work we need to exploit the inner structure of Razborov's proof. More specifically, we need to observe that Razborov's construction can be applied `inside' the oracle circuit so that we can replace each oracle by the majority of its random approximators and still make the case analysis work. This requires one more modification: we need to use Ajtai's p-size constant-depth circuits for the approximate majority instead of general circuits \cite{Afla}.

\subsection{Related work}

Our results are motivated primarily by the development of hardness magnification and the approximation method, but the notion of locality was considered already by Yao \cite{Yl}, who observed that some concrete monotone circuit lower bounds such as the lower bound of Razborov \cite{Rmclb1} localize. Yao's observations are incomparable to our results: Our results are not restricted to specific instantiations of the approximation method, but they work only for non-monotone circuits.

\subsection{Open problems}

\noindent {\bf Optimal localizability.} Do superlinear lower bounds for general circuits based on the approximation method localize even with a linear number of oracles (cf. Problem \ref{pr:fullloc})? Regarding constant-depth circuits, Theorem \ref{t:idloc} localizes lower bounds for depth $d$ circuits of size $poly(n2^{8m/d})$. In the HM frontier in the introduction, Item 1 uses $m=poly(\log n)$ while the lower bound in Item 3 is just polynomial. Is it possible to localize lower bounds for depth $d$ circuits of size $poly(n2^{m^{O(1/d)}})$? A stronger localizability of this kind could follow from a positive answer to Problem~\ref{pr:nc}, which asks for a barrier for deriving \NC lower bounds via the approximation method. It would be interesting to obtain such localizations even under additional assumptions of a constructivity of approximation models, e.g. assuming that the lower bounds based on these models are provable in \SB, a theory of bounded arithmetic formalizing p-time reasoning \cite{Bba}.
\bigskip

\noindent {\bf Random restrictions as the 
approximation method.} 
Is it possible to show that the approximation method is complete for constant-depth circuit lower bounds? Can we show that the method of random restrictions is formalizable in the framework of the approximation method? In both cases, we allow using inessential variables. Theorem \ref{t:dfcomp} suggests that such models should exist, but we have not managed to construct them. If we obtained such models with $O(\log s)$ inessential variables, they would be subject to the localizability from Theorem \ref{t:idloc} (with the asymmetric version of $\rho_d$). 
This would extend the generality of the approximation method and the impact of the localizability from Theorem \ref{t:idloc}. In particular, it would provide a unifying explanation for the localizability of both Razborov-Smolensky lower bound method and the method of random restrictions.
\bigskip

\noindent {\bf Feasible and structured models.} 
Razborov \cite{Ram2} introduced a generalization of the approximation method which can prove $s$-size lower bounds for almost all functions $f$ with error sets of size $poly(s)$. This required detaching errors sets from the inputs of $f$ so that the earlier barrier result would not apply. A drawback of the generalized approximation method is that it is not known to be complete and the models from \cite{Ram2} were obtained nonconstructively. 
Nevertheless, the elegance and restrictiveness of Razborov's generalized approximation models could be instrumental when searching for concrete examples of approximation models. It could thus be useful to find a constructive version of Razborov's models. Here, constructivity means the existence of a p-time algorithm which given truth-tables of functions $f_1,f_2$ of the model, outputs the truth-table of the `approximated' conjuction/disjunction of $f_1$ and $f_2$. This question was posed already in \cite{Ram2}. In order for the approximation models obtained from the fusion method for partial functions to be constructive, we need efficient algorithms generating anticheckers. Can we obtain such models unconditionally? 
\bigskip

\noindent {\bf Nonlocalizable lower bounds.} I believe that Barrington-Straubing \cite{BS} lower bound for $\Omega(n\log\log n)$-size formulas of constant-depth with parity gates can be used to show the hardness of $\MCSP[n^c,2n^{c}]$ for $\Omega(n\log\log n)$-size formulas of constant-depth with parity gates at the bottom. 
 Since $\MCSP[n^c,2n^{c}]$ is computable by $\AC$-{\sf XOR}$[O(N)]$ with local oracles, if we could make the lower bound work for circuits instead of formulas, it would give us an example of a concrete nonlocalizable lower bound. There are several other lower bounds which work only for computational models of nearly linear size. Is it possible that the weakness of these lower bounds is a result of their sensitivity to the presence of local oracles? If so, their `weakness' would turn out to be their strength. 

\section{Preliminaries}

$[n]$ denotes $\{1,\dots,n\}$. 
Let ${\cal F}_n$ be the set of all Boolean functions on $n$ inputs. 
We identify a Boolean function $f\in {\cal F}_n$ with the set of its ones $\{x\mid f(x)=1\}$.

A Boolean circuit with inputs $x_1,\dots, x_n$ is a directed acyclic graph (without multi-edges) 
 with vertices labeled by inputs $x_1,\dots, x_n$, constants 0,1 or commutative\footnote{As the underlying acyclic graph of a circuit does not specify the order of inputs to a gate, for the computation of the circuit to be well-defined, we need the connectives to be commutative. That is, $\circ(y_1,\dots,y_m)=\circ(y_{\pi(1)},\dots,y_{\pi(m)})$ for each permutation $\pi$ on $[m]$.} connectives $\circ(y_1,\dots,y_m):\{0,1\}^m\mapsto \{0,1\}$ so that there are exactly $m$ edges directed to the vertex labeled by $\circ(y_1,\dots,y_m)$. Non-input vertices are called gates. One of the gates of the circuit is designed as the output gate. A Boolean circuit thus computes a Boolean function. We consider only Boolean circuits with a single output gate. The depth of a circuit is the number of edges on 
 the longest path from the output to an input or to a constant. The size of a circuit is the number of its gates.

A Boolean formula is a Boolean circuit with a tree as the underlying graph and with logical connectives $\neg$, binary $\wedge$, binary $\vee$. We make an exception and measure the size of a Boolean formula by the number of leafs of its underlying tree.

A circuit class $\cal C$ is just a set of circuits - we do not impose any `composition' properties on it. $\circ(\cal C)$ denotes the set of connectives of circuits from $\cal C$. For example, for a class $\cal C$ of constant-depth circuits, $\circ({\cal C})$ can be the set of logical connectives of unbounded arity which is, formally, defined as the set of connectives $\neg, \vee_m,\wedge_m$, for each possible arity $m\ge 2$. We say that circuits from $\cal C$ are `Boolean circuits over $\circ({\cal C})$' and `$\cal C$-circuits'.

$\Circuit[s]$ denotes Boolean circuits over $\{\neg,\wedge_2,\vee_2\}$ of size at most $s$.
Given $f\in {\cal F}_n$, $Size_{{\cal C}}(f)$ denotes the size of the smallest $\cal C$-circuit computing $f$, if one exists. If there is no such circuit, we let $Size_{{\cal C}}(f):=\infty$.

\subsection{Approximation models}\label{s:models}

Intuitively, the approximation method formalizes the following strategy for proving circuit lower bounds: First, show that each $s$-size circuit $C$ can be approximated by an $s$-size circuit $\overline{C}$ so that $\Pr_x[C(x)\ne \overline{C}(x)]$ is small. The approximating circuit $\overline{C}$ is obtained from $C$ by replacing each gate of $C$ by an approximating gate. Then show that each $s$-size approximating circuit $\overline{C}$ is far from $f$ in the sense that $\Pr_x[\overline{C}(x)=f(x)]$ is small. Consequently, no $s$-size circuit $C$ computes $f$ either. We proceed with formal definitions. 

\begin{definition}[Legitimate model]\label{d:legmodel} A legitimate model (of order $n$) of a circuit class ${\cal C}$ is a pair $\left<{\cal M}, \overline{\circ}({\cal M})\right>$, where $\{0,1, x_i \mid i\ge 1 \}\subseteq {\cal M}\subseteq {\cal F}_n$ and $\overline{\circ}({\cal M})$ is a set of commutative operations $${\overline \circ}:\underbrace{{\cal M}\times\dots \times {\cal M}}_\text{$m$ times}\mapsto {\cal M},$$ for each $\circ\in \circ({\cal C})$ of arity $m$. We abuse the notation and denote $\left<{\cal M}, \overline{\circ}({\cal M})\right>$ by $\cal M$.
\end{definition}

Note that the `approximator' $\overline{\circ}$ of $\circ\in\circ({\cal C})$ depends on the input functions. This means that a single connective of a circuit can be approximated by different functions in different positions in the circuit.

\begin{definition}[Approximating circuit]
Given a legitimate model (of order $n$) of a circuit class ${\cal C}$ and a 
 $\cal C$-circuit $C$ with inputs from $\{x_1,\dots,x_n\}$, we define an approximating circuit $\overline{C}$ inductively by replacing each gate $\circ (C_1,\dots,C_m)$ in $C$ by $\overline{\circ}([[\overline{C_1}]],\dots,[[\overline{C_m}]])$ and setting $\overline{1}:=1, \overline{0}:=0, \overline{x_i}:=x_i$. Here, $[[B]]\in {\cal F}_n$ denotes the function computed by circuit (or approximating circuit) $B$. In order to simplify the notation, we will assume that $B$ is $[[B]]$, if it is used in a context which asks for a function.
\end{definition}

\begin{definition}[Error sets]\label{d:errors} Given a legitimate model $\cal M$ of a circuit class ${\cal C}$, we define the error sets of $\cal M$ by \begin{align*}\delta^+_{\circ}(f_1,\dots,f_m):= & \circ(f_1,\dots,f_m)\backslash \overline{\circ}(f_1,\dots,f_m), \\
\delta^-_{\circ}(f_1,\dots,f_m):= &\ \overline{\circ}(f_1,\dots,f_m)\backslash \circ(f_1,\dots,f_m),\\
\delta_{\circ}(f_1,\dots,f_m):= &\ \delta^{+}_{\circ}(f_1,\dots,f_m)\cup \delta^-_{\circ}(f_1,\dots,f_m),
\end{align*}
where $\circ\in \circ({\cal C})$ has arity $m$ and $f_1,\dots, f_m\in \cal M$. Further, we let $$\Delta:= \{\delta_{\circ}(f_1,\dots,f_m)\mid f_1,\dots,f_m\in {\cal M}, \circ\in\circ({\cal C})\text{ of arity }m\}.$$
\end{definition}

\begin{definition}[Distance]\label{d:distance} 
 The distance $\rho(f,\cal M)$ of a legitimate model $\cal M$ of a circuit class $\cal C$ from a Boolean function $f\in {\cal F}_n$ is the minimal $t$ such that there are tuples $\left< \overline{\circ_i}, f^i_1,\dots, f^i_{m_i} \right>$, where $m_i$ is the arity of $\circ_i\in\circ({\cal C})$, $i=1,\dots, t$ and $f^i_j\in \cal M$, such that for some $g\in\cal M$, 
$$f\oplus g\subseteq\  \bigcup_{i=1,\dots,t} \delta_{\circ_i}(f_1^i,\dots,f_{m_i}^i).$$
If there is no such $t$, we set $\rho(f,{\cal M}):=\infty$.\footnote{
 Razborov \cite{Ram1} defined the distance $\rho(f,\cal M)$ as the minimal $t$ such that $$f\backslash g\subseteq\  \bigcup_{i=1,\dots,t} \delta^+_{\circ_i}(f_1^i,\dots,f_{m_i}^i) \ \ \text{and} \ \ 
g\backslash f \subseteq\ \bigcup_{i=1,\dots, t} \delta^-_{\circ_i}(f_1^i,\dots,f_{m_i}^i).$$ We use a `symmetric' definition of $\rho$, which simplifies proofs, but our results work for Razborov's `asymmetric' $\rho$ as well (with an appropriate adjustment of parameters). The asymmetry in Razborov's definition of $\rho$ is used when defining the fusion method, see \S\ref{s:fusion}.}
\end{definition}

\begin{proposition}\label{p:approxbound} For each $f\in {\cal F}_n$ and each legitimate model ${\cal M}$ of a circuit class $\cal C$, 
$$\rho(f,{\cal M})\le Size_{{\cal C}}(f).$$
\end{proposition}

\proof
Let $C$ be a $\cal C$-circuit computing $f$ and $g:=\overline{C}\in {\cal F}_n$. 
Consider the set of tuples $\left< \overline{\circ_i}, \overline{B_1},\dots,\overline{B_{m_i}}\right>$ for all gates $\circ_i$ of $C$ together with the subcircuits $B_1,\dots,B_{m_i}$ which act as inputs of $\circ_i$. If $f(x)\ne g(x)$, we can compare computations of $C$ and $\overline{C}$ on $x$. We start at the output gate of $C$ and proceed to some input of the output gate which preserves the error, if such an input exists. We proceed in this way until we reach a gate $\circ$ of $C$ such that $\circ(D_1(x),\dots,D_{m}(x))\ne \overline{\circ}(\overline{D_1}(x),\dots,\overline{D_{m}}(x))$ but $D_j(x)=\overline{D_j}(x)$, for subcircuits $D_1,\dots,D_{m}$ acting as inputs of $\circ$. Therefore, $x\in \delta_{\circ}(\overline{D_1},\dots,\overline{D_{m}})$. 
\qed
\bigskip

\noindent{{\bf Probabilistic approach.}} A particularly useful instantiation of the approximation method can be described as the following probabilistic approach. For a random variable ${\bm x}$ on $\{0,1\}^n$, let $$d:=\max \Big\{\Pr[{\bm x}\in \delta_{\circ}(f_1,\dots,f_m)] \mid f_1,\dots, f_m\in {\cal M}, \circ\in\circ({\cal C})\text{ of arity }m\Big\}.$$

If $d\ne 0$, define $$\rho(f,{\cal M},{\bm x}):=\min_{g\in{\cal M}} \Bigg\{\frac{\Pr[f({\bm x})\ne g({\bm x})]}{d}\Bigg\}.$$ 
Assuming $\rho(f,{\cal M})\le t$ is witnessed by $g$ and a set $T$ of error sets with $|T|=t$, we have $\Pr[f({\bm x})\ne g({\bm x})]\le \sum_{\delta\in T}\Pr[{\bm x}\in \delta]\le dt$. Therefore, $\rho(f,{\cal M},{\bm x})\le \rho(f,{\cal M})$. 
\medskip

\noindent{{\bf Examples.}} Lower bounds of Razborov-Smolensky for $\AC[p]$ and monotone circuit lower bounds of Razborov can be naturally formulated in the framework of the approximation method. In fact, they use the probabilistic approach described above. 
\smallskip

\noindent {\em Monotone circuit lower bounds:} We refer to the presentation in Arora-Barak \cite[Chapter 14.3]{AB}. The approximation model ${\cal M}$ consists of functions computed by small disjunctions of indicator functions. The operations $\overline{\vee},\overline{\wedge}$ are defined so that probabilities $\Pr[{\bm x}\in \delta_{\circ}]$ are small, for a suitable ${\bm x}$. Nevertheless, for each $g\in\cal M$, $\Pr[f({\bm x})\ne g({\bm x})]$ is high, if $f$ is a suitable function, so $\rho(f,{\cal M},{\bm x})$ yields the desired lower bound.
\smallskip

\noindent {\em $\AC[p]$ lower bounds:} We refer to the presentation in \cite{MP} or \cite[Chapter 14.2]{AB}. 
The lower bound for $\AC[p]$ circuits of depth $d$ is obtained by a lower bound for a version of $\rho(f,{\cal M},{\bm x})$, where we consider the minimum only over functions $g\in {\cal M}$ such that for some circuit $C$ of depth $d$, $g=\overline{C}$. 
 The approximation model $\cal M$ consists of functions computed by polynomials over a finite field. The operations $\neg, MOD_p$ are approximated by polynomials which make no errors and the operations $\wedge_m,\vee_m$ (of unbounded arity)  are approximated by polynomials which err only on a small number of inputs (The approximating polynomials actually depend on the position of the gate in the given circuit.). On the other hand, functions $g$ such that $g=\overline{C}$, for a circuit of depth $d$, are low-degree polynomials and the probability that a low-degree polynomial errs to compute $MOD_q$, for $q\ne p$, is high.

\section{Limitations}
\subsection{General circuits}

Razborov \cite{Ram1} showed inherent limitations of the approximation method, see also \cite{Olf}. 
 We present Razborov's proof and then adapt it to derive localizations of lower bounds based on the approximation method.

\begin{theorem}[Razborov \cite{Ram1}]\label{t:amlim} For each legitimate model $\cal M$ of Boolean circuits over $\{\neg,\vee_2,\wedge_2\}$ such that $\overline{\neg}:=\neg$, and for each $f\in {\cal F}_n$, $$\rho(f,{\cal M})\le O(n_0n),$$ where $n_0$ is the number of essential inputs of $f$, i.e. inputs of $f$ such that switching the value of the input affects the output of $f$ on some assignment of the remaining inputs.  Moreover, for every random variable $\bm x$ on $\{0,1\}^n$, $$\rho(f,{\cal M}, {\bm x})\le 12(n_0+1).$$
\end{theorem}

\proof We start with the `moreover' part. For each legitimate model $\cal M$, a Boolean function $f$ and a random variable ${\bm x}$ on $\{0,1\}^n$, we want to find a function $g\in\cal M$ such that $\Pr[g({\bm x})\ne f({\bm x})]\le 12(n_0+1)d$. We use bold font to denote random variables. The existence of such $g$ follows from the following inequality, $$\Pr[{\bm h}({\bm x})\ne f({\bm x})]\le 12(n_0+1)\Pr[{\bm x}\in {\bm \delta}]\le 12(n_0+1)d.$$ Here, the second inequality holds by the definition of $d$. The first inequality is a direct corollary of Lemma \ref{l:distr}.

\begin{lemma}\label{l:distr} For each legitimate model $\cal M$ of Boolean circuits over $\{\vee_2,\wedge_2,\neg\}$ such that $\overline{\neg}:=\neg$, and for each $f\in {\cal F}_n$, there is a random variable ${\bm \delta}$ on $\Delta$ and a random variable $\bm h$ on $\cal M$ such that for each $x$, $$\Pr[{\bm h}(x)\ne f(x)]\le 12(n_0+1) \Pr[x\in \bm{\delta}].$$
\end{lemma}

Before proving Lemma \ref{l:distr} we show how it implies the rest of the theorem. Given a legitimate model $\cal M$ and a Boolean function $f$, we want to find a function $g\in\cal M$ and $O(nn_0)$ tuples $\langle \overline{\circ_i},f_1^i,\dots,f_{m_i}^i\rangle$ which cover $f\oplus g$. Observe that there are two ways of `upper-bounding' $\rho(f,{\cal M})$. 
\smallskip

\noindent {\bf Case I (random approximators).} Consider inputs $x$ such that $\Pr[{\bm h}(x)\ne f(x)]<1/3$. 
By (the multiplicative) 
 Chernoff's bound, $$\Pr[\mathsf{MAJ}({\bm h}_1,\dots, {\bm h}_k)(x)\ne f(x)]\le 1/e^{k/48},$$ where each ${\bm h}_i$ is an independent copy of $\bm h$. Consequently, we can fix $h_1,\dots,h_{O(n)}$ such that $\mathsf{MAJ}(h_1,\dots,h_{O(n)})$ 
 coincides with $f$ on all such inputs $x$. As the majority function {\sf MAJ} on $m$ inputs is computable 
 by an $O(m\log m)$-size circuit\footnote{Razborov uses monotone $O(m\log m)$-size circuits for $\mathsf{MAJ}$, which is needed to make the proof work for his definition of $\rho$.} $M$, we set $g:=\overline{M}(h_1,\dots,h_{O(n)})$ and 
 conclude that all inputs $x\in g\oplus f$ 
 considered in Case I are covered by $O(n\log n)\le O(nn_0)$ tuples. Here, w.l.o.g. $n_0\ge \log n$ since otherwise $\rho(f,{\cal M})\le 2^{n_0}\le n$.
\smallskip

\noindent {\bf Case II (random error sets).} For inputs $x$ such that $\Pr[{\bm h}(x)\ne f(x)]\ge 1/3$, we use the fact that random $\delta$ covers such $x$ with high probability. That is, by Lemma \ref{l:distr}, $$\Pr[x\in{\bm \delta}_1\vee\dots\vee x\in {\bm \delta}_{k'}]\ge 1-1/e^{k'/36(n_0+1)},$$ where each ${\bm \delta}_i$ is an independent copy of $\bm \delta$. Therefore, there are $\delta_1,\dots,\delta_{O(nn_0)}$ covering all $x$'s considered in Case II.
\smallskip

Cases I and II together imply the inequality $\rho(f,{\cal M})\le O(nn_0)$. 
\bigskip

It remains to prove Lemma \ref{l:distr}.

Given $h\in {\cal F}_m$, let $h^{a}:=h(x_1,\dots,x_{m-1},a)\in {\cal F}_{m-1}$, for $a\in\{0,1\}$. Further, let $C_h$ be a trivial exponential-size circuit computing $h$ defined inductively as $$C_h:=(C_{h^1}\wedge x_m)\vee (C_{h^0}\wedge \neg x_m),$$ with $C_{h(b)}:=h(b)$, for $b\in \{0,1\}^m$. Define the random variable $\bm h:=\overline{D_{{\bm g}}}$, where $$D_{{\bm g}}:=(C_{f\oplus {\bm g}}\wedge C_{\neg {\bm g}})\vee (C_{\neg (f\oplus {\bm g})}\wedge C_{{\bm g}}),$$ for uniformly random ${\bm g}$ on ${\cal F}_n$. Note that for each $g\in {\cal F}_n$, 
$D_g(x)=(f\oplus g\oplus g)(x)=f(x)$.

The random variable $\bm \delta$ is defined as a random error set of $\bm h$ as follows. Let $m\in [n]\cup\{\oplus\}, t\in \{0,1,\vee\}$ and ${\bm g_m}\in{\cal F}_m$, with ${\bm g_{\oplus}}:={\bm g_n}\in{\cal F}_{\oplus}:={\cal F}_n$, 
 be chosen independently and uniformly at random.

\[{\bm \delta}:=
\begin{cases}
\delta_{\vee}(\overline{C_{f\oplus {\bm g_m}}\wedge C_{\neg {\bm g_m}}}, \overline{C_{\neg (f\oplus {\bm g_m})}\wedge C_{{\bm g_m}}}) & \text{if } m=\oplus, t=\vee,\\
\delta_{\wedge}(\overline{C_{t\oplus f\oplus {\bm g_m}}}, \overline{C_{t\oplus \neg {\bm g_m}}}) & \text{if } m=\oplus, t\in \{0,1\},\\
\delta_{\vee}(\overline{C_{{\bm g_{m}}^1}\wedge x_m}, \overline{C_{{\bm g_{m}}^0}\wedge\neg x_m}) & \text{if } m\in [n], t=\vee,\\
\delta_{\wedge}(\overline{C_{{\bm g_{m}}^{1-t}}}, \overline{ \underbrace{\neg}_\text{$t$ times} x_m}) & \text{if } m\in [n], t=\{0,1\}.
\end{cases}
\]

The crucial observation is provided by Claim \ref{cl:types}.

\begin{claim}\label{cl:types} For each $x$, there are $3+12n$ positions of gates in $D_{g}$ (the underlying graph of $D_g$ is independent of $g$) such that for each $g\in {\cal F}_n$ satisfying $\overline{D_g}(x)\ne f(x)$, we have $x\in \delta_{\circ}(\overline{e_1},\overline{e_2})$, where $\circ$ is the gate of $D_g$ corresponding to one of these $3+12n$ positions and $e_1,e_2$ are its inputs. 
\end{claim}

To prove Claim \ref{cl:types}, note that the error can occur either in one of the top 3 gates of $D_g$ or in one of the remaining 4 subcircuits. Consider $C_g$, other circuits are treated analogously. Suppose $x$ is a `+error' in the sense that $C_g(x)>\overline{C_g}(x)$. Then, there are gates $e_1,e_2,e_3$ in $C_g$ such that $e_3=e_1\circ e_2$ and $\overline{e_3}$ makes a +error on $x$, but $\overline{e_1}, \overline{e_2}$ do not, i.e. $e_3(x)>\overline{e_3}(x)$, $e_1(x)\le\overline{e_1}(x)$, $e_2(x)\le\overline{e_2}(x)$.\footnote{Gate $e_i$ computes the function defined by the corresponding subcircuit of $C_g$.} 
 Moreover, as one of the conjunctions $C_{g^0}(x)\wedge x_m$, $C_{g^0}(x)\wedge \neg x_m$ is false, a +error occurs in some of $2n$ gates on the path through $C_g$ which is consistent with $x$. If a +error occurs in $\overline{e_3}$, but not in $\overline{e_1},\overline{e_2}$, then $x\in\delta^+_{\circ}(\overline{e_1},\overline{e_2})$ because $$\overline{e_1}(x)\circ \overline{e_2}(x)\ge e_1(x)\circ e_2(x)=e_3(x)>\overline{e_3}(x)=\overline{e_1}(x){\overline\circ}\ \overline{e_2}(x).$$ If $x$ is a `--error' satisfying $C_g(x)<\overline{C_g}(x)$, we can again find gates $e_1,e_2,e_3$ in $C_g$ such that $e_3=e_1\circ e_2$ and $\overline{e_3}$ makes a --error on $x$ but $\overline{e_1},\overline{e_2}$ do not. Analogously as before, this implies $x\in \delta^-_{\circ}(\overline{e_1},\overline{e_2})$.  If there is a -error on $C_h(x)\wedge \neg x_m$ which is inconsistent with $x$, then $\overline{\neg x_m}(x)=\neg x_m=0$ and we conclude $x\in\delta^-_{\circ}(\overline{e_1},\overline{e_2})$, for $e_1=C_{h}$ and $e_2=\neg x_m$, 
 as $$\overline{e_1}(x)\overline{\wedge}\overline{e_2}(x)=\overline{e_3}(x)>e_3(x)\ge \overline{e_1}(x)\wedge \overline{e_2}(x).$$ 
 Therefore, a --error is covered by an error set corresponding to one of $3n$ positions in $C_g$: $2n$ positions on the path consistent with $x$ and $n$ positions deviating from the path. (The
 extra $n$ results from the fact that both disjuncts of $C_h$ can make a -error, but whenever a --error deviates from the path consistent with $x$ it is covered immediately on the first gate on which it deviates.) This proves the claim.
\smallskip

Having Claim \ref{cl:types}, we can finish the proof of Lemma \ref{l:distr}. For each $\delta$ in the range of ${\bm \delta}$, let $T(\delta):=\left<m,t\right>$, for $m\in [n]\cup\{\oplus\}, t\in \{0,1,\vee\}$ given by $\delta$, be the `type' of $\delta$. That is, there are $3(n+1)$ possible types. 
 Similarly, for fixed $x$, $g\in {\cal F}_n$, and $h=\overline{D_g}$ such that $h(x)\ne f(x)$, let $T(h):=T(\delta_{\circ}(\overline{e_1},\overline{e_2}))$ for $\delta_{\circ}(\overline{e_1},\overline{e_2})$ corresponding to the first of $3+12n$ positions from Claim \ref{cl:types} covering $x$ - note that $\delta_{\circ}(\overline{e_1},\overline{e_2})$ determines $m\in [n]\cup\{\oplus\}, t\in \{0,1,\vee\}, g_m\in {\cal F}_m$ 
 and the corresponding $\delta$ in the range of $\bm \delta$ such that $\delta_{\circ}(\overline{e_1},\overline{e_2})=\delta$. We want to show that 
\begin{align*}\Pr[x\in {\bm \delta}] & \ge\sum_{k=\left<m,t\right>} \frac{\Pr[x\in {\bm\delta} \mid T({\bm \delta})=k]}{3(n+1)}\\ & \ge \sum_{k=\left<m,t\right>} \frac{\Pr[{\bm h}(x)\ne f(x)\wedge T({\bm h})=k]}{12(n+1)}=\frac{\Pr[{\bm h}(x)\ne f(x)]}{12(n+1)}.\end{align*} It thus remains to prove 
 that $\Pr[x\in {\bm \delta}\mid T({\bm\delta})=k]\ge \Pr[{\bm h}(x)\ne f(x)\wedge T({\bm h})=k]/4$. For $m$ given by type $k$, we have $\Pr[x\in {\bm \delta}\mid T({\bm \delta})=k]=\Pr_{g_m\in {\cal F}_m}[x\in \delta^1]$, where $\delta^1$ is the error set from the range of ${\bm \delta}$ determined by $k$ and $g_m$. Further, $\Pr[{\bm h}(x)\ne f(x)\wedge T({\bm h})=k]=\Pr_{g_n\in {\cal F}_n}[\overline{D_{g_n}}(x)\ne f(x)\wedge T(\overline{D_{g_n}})=k]\le \Pr_{g_n\in {\cal F}_n}[\bigvee_i x\in \delta^2_i]$ where $\delta^2_i$ corresponds to the $i$th position of type $k$ from $3+12n$ positions given by Claim \ref{cl:types} (i.e. $1\le i\le 4 $). Finally, since all functions $f\oplus {\bm g},\neg {\bm g}, \neg (f\oplus {\bm g}),\bm g$ are uniformly random, $\Pr_{g_n\in {\cal F}_n}[\bigvee_i x\in \delta^2_i]\le 4 \Pr_{g_m\in {\cal F}_m}[x\in \delta^1]$.
  \qed

 \bigskip

The barrier from Theorem \ref{t:amlim} can be strengthened if we assume that approximating circuits $\overline{C}$ depend only on the essential variables of $C$. We refer to approximation models with this property as 0-projective models. (In Theorem \ref{t:genloc} we generalize 0-projectivity to `full' projectivity.)

\begin{definition}[0-projective model]
Let $\cal M$ be a legitimate model of a circuit class $\cal C$. We say that $\cal M$ is 0-projective if the condition $*$ holds:
\begin{itemize}
\item[$*$] If $C$ is a $\cal C$-circuit with inputs $x_j$, for $j\in J\subseteq \{0,1\}^n$, then for each $z,y\in\{0,1\}^n$, $$z|_J=y|_J\quad \Rightarrow\quad \overline{C}(z)=\overline{C}(y).$$ Here, $z|_J\in\{0,1\}^{|J|}$ is a projection of $z$ to bits $z_j$ with
 $j\in J$. 
\end{itemize}
\end{definition}

Lower bounds of Razborov-Smolensky for $\AC[p]$ and monotone circuit lower bounds of Razborov are obtained by constructing 0-projective approximation models. 

\begin{corollary}[A limitation of the 0-projective approximation method]\label{c:dim} Let $\cal M$ be any 0-projective legitimate model of Boolean circuits over $\{\neg, \wedge_2,\vee_2\}$ such that $\overline{\neg}:=\neg$. Then, for each $f\in {\cal F}_n$ with $n_0$ essential variables, $$\rho(f,{\cal M})\le O(n_0^2).$$
\end{corollary}

\proof We proceed as in the proof of Theorem \ref{t:amlim} but note that for 0-projective approximation models, for each $x,y\in \{0,1\}^n$ and error set $\delta_{\circ}(\overline{C_1},\overline{C_2})$, where $C_1,C_2$ have inputs from  $\{x_j \mid j\in J\subseteq \{0,1\}^n\}$, we have that \begin{equation}\label{e:dim}x\in \delta_{\circ}(\overline{C_1},\overline{C_2})\wedge x|_J=y|_J\quad\Rightarrow\quad y\in \delta_{\circ}(\overline{C_1},\overline{C_2}).
\end{equation} Further, by 0-projectivity of $\cal M$, each $h$ in the range of ${\bm h}$ from Lemma \ref{l:distr} depends just on the essential variables of $f$. 
Hence, in Case I it suffices to cover projections of $x$ on the essential variables, which can be done by $O(n_0\log n_0)$ tuples. Similarly, each $\delta$ from the range of 
 ${\bm \delta}$ in Lemma $\ref{l:distr}$ is of the form $\delta_{\circ}(\overline{C_1},\overline{C_2})$, where $C_1,C_2$ depend only on the essential variables of $f$. By (\ref{e:dim}), in Case II it thus again suffices to cover projections of $x$ on the essential variables, which can be done by $O(n_0^2)$ tuples. \qed
\bigskip

Theorem \ref{t:amlim} and Corollary \ref{c:dim} leave open the possibility of proving a superpolynomial circuit lower bound by proving a superlinear lower bound using standard or 
 0-projective approximation method and combining it with hardness magnification. We rule out such attacks for localizable magnification theorems under 
additional assumptions: 1. There is 
 only a sublinear number of oracles (in the oracle circuit implying the magnification theorem); 2. Assuming 0-projectivity, we allow a superlinear number of oracles, but they have to appear only at the bottom of the circuit; 3. Assuming a strengthening of 0-projectivity, we obtain the ideal localizabilty.

\begin{theorem}[Localizability of the approximation method for general circuits]\label{t:genloc} Let $\cal M$ be any legitimate model $\cal M$ of Boolean circuits over $\{\neg, \wedge_2,\vee_2\}$ such that $\overline{\neg}:=\neg$, and $f\in {\cal F}_n$. 
\begin{itemize}
\item[1.] {\em [Localizability with few oracles]} Suppose that $\rho(f,{\cal M})\ge s$. Then, $f$ is not computable by circuits of size $s-O(knm)$ using connectives $\neg,\wedge_2,\vee_2$ and $k$ arbitrarily powerful, possibly different and noncommutative\footnote{Allowing noncommutative oracle requires generalizing the notion of circuit so that its underlying graphs specifies the order of inputs.}, oracles of arity $m$.\footnote{It is possible to prove Item 1 with Razborov's asymmetric definition of $\rho(f,{\cal M})$, if we adjust the resulting lower bound for the oracle circuits to $s-O(kn\cdot max\{m,\log n\})$ and conclude that it holds only for circuits with monotone oracles and negations at the bottom. 
The term $max\{m,\log n\}$ comes from not using DNFs in the proof, which is done in order to keep negations at the bottom. Further, it is possible to avoid the requirement on the monotonicity of oracles and negations at the bottom by using slice functions (and adjusting the parameters appropriately).  
}

\item[2.] {\em [Localizability of 0-projective models]} Suppose that $\rho(f,{\cal M})\ge s$ and that $\cal M$ is 0-projective. Then, $f$ is not computable by circuits of size $s-O(km^2)$ using connectives $\neg,\wedge_2,\vee_2$ and $k$ arbitrarily powerful, possibly different and noncommutative, oracles of arity $m$ whose inputs are among $\{x_1,\neg x_1,\dots, x_n,\neg x_n\}$.

\item[3.] {\em [Localizability of projective models]} Suppose that $\rho(f,{\cal M})\ge s$ and that ${\cal M}$ is projective. Here, $\cal M$ is projective if for each Boolean circuit $C$ over $\{\neg,\wedge_2,\vee_2\}$ with $m$ inputs $x_1,\dots,x_m$, for all $f_1,\dots,f_m\in {\cal M}$ and each $z,y\in\{0,1\}^n$, $$\bigwedge_{i=1,\dots,m}f_i(z)=f_i(y)\quad \Rightarrow\quad \overline{C}(f_1(z),\dots,f_m(z))=\overline{C}(f_1(y),\dots,f_m(y)).$$ Then, $f$ is not computable by circuits of size $s-O(km^2)$ using connectives $\neg,\wedge_2,\vee_2$ and $k$ arbitrarily powerful, possibly different and noncommutative, oracles of arity $m$. 
\end{itemize}
\end{theorem}

While Theorem \ref{t:genloc} 
 is stated for general parameters, it is interesting only for $s<Kn^2$, where $K$ is the constant implicit in Theorem \ref{t:amlim}. Typical parameters of our interest are: $s=n^{1.99}, k=n^{1.88}, m=poly(\log n)$. Item 1 of Theorem \ref{t:genloc} is thus not satisfactory because it requires  $k<Kn/m$ for a nontrivial conclusion. 
 Item 2 works for the parameters of our interest but it requires oracles at the bottom (which is not the case for the HM frontier in the introduction and many other magnification theorems). By strengthening 0-projectivity to projective models, 
 in Item 3 we can allow oracles to appear anywhere in the circuit. Razborov-Smolensky lower bound method is projective, but Razborov's lower bound for monotone circuits is not. However, for monotone circuit lower bounds already Theorem \ref{t:amlim} does not apply.

\proof We start with the proof of Item 1 and then adapt it to obtain Items 2 and 3. 
\bigskip

\noindent Item 1: We prove the contrapositive. 
 Assume $f$ is computable by an $s$-size oracle circuit $C^O$ and that the gates of $C^O$ are ordered so that inputs of the $i$th gate $e^i$ precede $e^i$. Following our earlier convention, $e^i$ computes the function defined by the corresponding subcircuit of $C^O$.  We will inductively associate each $e^i$ with a function $f^i\in {\cal M}$ so that each $x$ satisfying $e^j(x)\ne f^j(x)$, for some $j\le i$, is covered by one of $i+Ko_imn$ error sets $\delta\in\Delta$, where $o_i$ is the number of oracles among the first $i$ gates of $C^O$ and $K$ is an absolute constant. In particular, we will have $\rho(e^i,{\cal M})\le i+Ko_imn$, $e^s=f$ and $o_s=k$.

Inputs $x_j$ and constants 0,1 are associated with functions $x_j,0,1\in {\cal M}$. 
 Suppose we already associated the first $i$ gates. If the $(i+1)$'st gate $\circ(e^{i_1},e^{i_2})$, with $i_1,i_2\le i$, computes $\neg,\wedge_2$ or $\vee_2$, we set $f^{i+1}:=\overline{\circ}(f^{i_1},f^{i_2})$. In this case we use one error set $\delta_{\circ}(f^{i_1},f^{i_2})$ to cover all $x$'s such that $\overline{\circ}(f^{i_1},f^{i_2})(x)\ne \circ(f^{i_1},f^{i_2})(x)$. By the inductive hypothesis, all $x$'s such that $f^{j}(x)\ne e^{j}(x)$, for some $j\le i$, are covered by $i+Ko_imn$ error sets, where $o_i=o_{i+1}$. Hence, all $x$'s such that $f^{j}(x)\ne e^{j}(x)$, for some $j\le i+1$, are covered by $i+1+Ko_{i+1}mn$ error sets.

Assume that the $(i+1)$st gate $\circ(e^{i_1},\dots,e^{i_m})$ computes an oracle of arity $m$. Following the proof of Lemma \ref{l:distr} (with $f^{i_j}$'s instead of $x_j$'s, with $\overline{f^{i_j}}:=f^{i_j}$ and $f=\circ$), we obtain a  random variable ${\bm\delta}$ on $\Delta$ and a random variable ${\bm h}$ on $\cal M$ such that for each $x$, $$\Pr[{\bm h}(x)\ne \circ(f^{i_1},\dots,f^{i_m})(x)]\le 12(m+1)\Pr[x\in {\bm \delta}].$$ Next, we simulate the case analysis from the proof of Theorem $\ref{t:amlim}$.

Case I: Consider $x$'s satisfying $\Pr[{\bm h}(x)\ne \circ(f^{i_1},\dots,f^{i_m})(x)]<1/3$. Let $M$ be a $K_1n\log n$-size circuit computing $\mathsf{MAJ}$ and $h_1,\dots,h_{K_2n}$ be such that on all $x$'s from Case I, $M(h_1(x),\dots,h_{K_2n}(x))=\circ(f^{i_1},\dots,f^{i_m})(x)$. Here, $K_1,K_2$ are absolute constants. That is, setting $f^{i+1}:=\overline{M}(h_1,\dots,h_{K_2n})$, there are $K_1n\log n$ error sets in $\Delta$ covering all $x$'s from Case I such that $f^{i+1}(x)\ne \circ(f^{i_1},\dots,f^{i_m})(x)$. W.l.o.g. $\log n\le m$, otherwise we can compute an oracle of arity $m$ by a circuit of size $O(n)$ 
 and use its approximator to define $f^{i+1}$ such that all errors of $f^{i+1}$ are covered by $O(n)$ error sets.

Case II: As in the proof of Theorem \ref{t:amlim}, the remaining $x$'s not considered in Case I are covered by $K_3mn$ error sets, for an absolute constant $K_3$. 

Cases I and II together with the inductive hypothesis imply that all $x$'s such that $f^{j}(x)\ne e^{j}(x)$, for some $j\le i+1$, are covered by $Ko_{i+1}mn+(i+1)$ 
 error sets. This finishes the proof of Item 1.
\bigskip

\noindent Item 2: We proceed as in the proof of Item 1 but assume additionally that oracles of $C^O$ are at the bottom. In fact, w.l.o.g. each oracle has the form $\circ(x_{i_1},\dots,x_{i_m})$ as otherwise (if some $x_j$ is $\neg x_j$) we can redefine $\circ$. When considering the $(i+1)$st oracle gate $\circ(e^{i_1},\dots,e^{i_m})$, by $*$-purity of $\cal M$, we conclude additionally that each $h$ in the range of $\bm h$ depends only on $x_{i_1},\dots,x_{i_m}$. This allows us to use just $K_2m$ functions $h$ as inputs of $M$ and $K_1m\log m$ covering error sets in Case I. In Case II, just like in the proof of Corollary \ref{c:dim}, we need only $O(m^2)$ error sets as well.
\bigskip

\noindent Item 3: Again, we follow the proof of Item 1 observing that for projective 
 models, for each $h$ in the range of $\bm h$, and each $x,y\in\{0,1\}^n$, $$\bigwedge_{j=1,\dots,m} f^{i_j}(x)=f^{i_j}(y)\quad \Rightarrow\quad h(x)=h(y).$$ 
Therefore, in Cases I and II it suffices to cover projections of $x$ on $f^{i_1}(x),\dots,f^{i_m}(x)$, which can be done with $O(m^2)$ error sets.
  \qed

\bigskip
\begin{problem}[Full localizability of the approximation method for general circuits]\label{pr:fullloc} Does Item 3 of Theorem \ref{t:genloc} hold without the assumption of projectivity?
\end{problem}

\bigskip
\noindent {\bf Models preserving the structure of circuits.} Proposition \ref{p:approxbound} holds even for a `structured'  $\rho(f,{\cal M})$ defined so that the tuples $\left< \overline{\circ}_i, f^i_1,\dots,f^i_{m_i}\right>$ are required to form a circuit. That is, the structured $\rho(f,{\cal M})$, denoted $\rho'(f,{\cal M})$, 
 is the minimal number of tuples covering $f\oplus g$, for some $g$, such that we can assign the tuples to nodes of a directed acyclic graph satisfying: 
\begin{itemize}
\item[1.] If nodes $\left< \overline{\circ}_{i_1}, f^{i_1}_1,\dots,f^{i_1}_{m_{i_1}}\right>,\dots, \left< \overline{\circ}_{i_k}, f^{i_k}_1,\dots,f^{i_k}_{m_{i_k}}\right>$ are inputs of node $\left< \overline{\circ}_{i}, f^{i}_1,\dots,f^i_{k}\right>$, then $f^{i}_j=\overline{\circ}_{i_j}(f^{i_j}_1,\dots,f^{i_j}_{m_{i_j}})$, for $j=1,\dots, k$;
\item[2.] For nodes $\left< \overline{\circ}_i, f^i_1,\dots,f^i_{m_i}\right>$ at the bottom, each $f^i_j$ is $x_k\in {\cal M}$ for some $k$. 
\end{itemize}
For $\rho'(f,{\cal M})$, the barrier from Theorem \ref{t:amlim} fails. Is it exactly the insensitivity to the structure of circuits what prevents 
 the approximation method from proving strong circuit lower bounds?
That is, is there a legitimate model $\cal M$ such that 
 $\rho'(f,{\cal M})\le s$ implies $f\in \Circuit[s^{O(1)}]$?

If we required that $g$ in the definition of $\rho(f,{\cal M})$ is $\overline{C}$ for a circuit $C$ of size $s$, the resulting method would become complete for $s$-size circuit lower bounds. This follows trivially by considering a model with empty error sets. 
\bigskip

\noindent {\bf Formula lower bounds via approximation models.}
 By imposing an additional structure of covering error sets in the definition of $\rho(f,{\cal M})$, it is possible to formulate versions of the approximation method which are, in principle, more suitable for proving formula lower bounds. One possibility is to require that the tuples from the definition of the structured $\rho(f,{\cal M})$ form not only a circuit but a formula. Is it possible to adapt Theorem \ref{t:amlim} to such formula-versions of $\rho(f,{\cal M})$? Note that the upper bounds from Theorem \ref{t:amlim} are witnessed by $g\in {\cal M}$ such that $g=\overline{C}$ for a formula $C$. In particular, circuits $D_{g}$ from the proof of Lemma \ref{l:distr}, which are used to define the covering error sets, are formulas. Obtaining such versions of Theorem \ref{t:amlim} for formulas would imply that known superquadratic formula lower bounds based on the method of random restrictions are not formalizable by the corresponding versions 
 of the approximation method.

\subsection{Constant-depth circuits}

\NC lower bounds can be approached also via the approximation method for constant-depth circuits. This is because p-size formulas are computable by depth $d+2$ 
 circuits of size $2^{n^{O(1/d)}}$. We use Theorem \ref{t:amlim} to impose a limitation of this approach. 

\begin{definition}[$\rho_d(f,\cal M)$]\label{d:ddistance} 
 The $d$-depth distance $\rho_d(f,\cal M)$ of a legitimate model $\cal M$ of a circuit class $\cal C$ from a Boolean function $f\in {\cal F}_n$ is the minimal $t$ such that there are tuples $\left< \overline{\circ_i}, f^i_1,\dots, f^i_{m_i} \right>$, where $m_i$ is the arity of $\circ_i\in\circ({\cal C})$, $i=1,\dots, t$ and $f^i_j\in \cal M$, such that for some $g\in\cal M$, $$f\oplus g\subseteq\  \bigcup_{i=1,\dots,t} \delta_{\circ_i}(f_1^i,\dots,f_{m_i}^i).$$ Moreover, for each $f^i_j$ above there is a $\cal C$-circuit $C$ of depth at most $d-1$ such that $f^i_j=\overline{C}$ and $g=\overline{D}$ for a $\cal C$-circuit $D$ of depth at most $d$. If there is no such $t$, set $\rho_{d}(f,{\cal M}):=\infty$.
\end{definition}

Proposition \ref{p:approxbound} can be adapted to the case of $\rho_d(f,{\cal M})$ showing that lower bounds for $\cal C$-circuits of depth $d$ can be obtained via lower bounds for $\rho_d(f,{\cal M})$. Razborov-Smolensky lower bound for $\AC[p]$ can be formulated as a lower bound for $\rho_d(f,{\cal M})$.
\smallskip

\begin{theorem}\label{t:damlim} For each legitimate model $\cal M$ of Boolean circuits over $\bigcup_{a\ge 2}\{\neg,\vee_a,\wedge_a\}$ such that $\overline{\neg}:=\neg$, for each $f\in {\cal F}_n$ and each $d\ge 8$, $$\rho_{d}(f,{\cal M})\le Knd(2^{\lceil n/\lfloor d/8\rfloor\rceil}+1)+2^{Kn^{k/d}}+Kn^K,$$ where $K,k$ are absolute constants (independent of $d$). 
\end{theorem}

\proof We follow the proof of Theorem \ref{t:amlim} and start by adapting Lemma \ref{l:distr}.

\begin{lemma}\label{l:ddistr} For each legitimate model $\cal M$ of Boolean circuits over $\bigcup_{a\ge 2}\{\vee_a,\wedge_a,\neg\}$ such that $\overline{\neg}:=\neg$, for each $f\in {\cal F}_n$ and each $d\ge 1$, there is a random variable ${\bm \delta}$ on $\Delta$ and a random variable $\bm h$ on $\cal M$ such that for each $x$, $$\Pr[{\bm h}(x)\ne f(x)]\le 4(d+1)(2^{\lceil n/d\rceil}+1)\Pr[x\in \bm{\delta}].$$ Moreover, for each $h$ in the range of ${\bm h}$ there is a circuit $D$ of depth $2d+3$ such that $h=\overline{D}$ and for each $\delta_{\circ}(f_1,\dots,f_m)$ in the range of $\bm \delta$ there are circuits $D_i$ of depth at most $2d+2$ such that $f_i=\overline{D_i}$. 
\end{lemma}

Lemma \ref{l:ddistr} implies Theorem
 \ref{t:damlim} by the following case analysis.
\smallskip

Case I: Consider inputs $x$ such that $\Pr[{\bm h}(x)\ne f(x)]<1/3$. Since $\mathsf{MAJ}$ on $m$ inputs is computable by a $poly(m)$-size formula, it is also computable by a depth $d+2$ 
 circuit $M$ of size $max\{2^{O(m^{k/d})},poly(m)\}$, for an absolute constant $k$. Hence, we can set $g:=\overline{M}(h_1,\dots,h_{O(n)})$ for suitable $h_1,\dots,h_{O(n)}\in {\cal M}$, so that each $x\in g\oplus f$ considered in Case I is covered by one of $2^{O(n^{k/d})}+poly(n)$ error sets. Note that $g$ is computed by an approximator of a circuit of depth $3d+5$. The error sets thus satisfy the `moreover' requirement from the definition of $\rho_{3d+5}$.

Case II: Similarly as in the proof of Theorem \ref{t:amlim}, the remaining inputs $x$ can be covered by $O((d+1)(2^{\lceil n/d\rceil}+1)n)$ error sets.

Cases I and II imply that $\rho_{8d}(f,{\cal M})\le \rho_{3d+5}(f,{\cal M})\le O(d(2^{\lceil n/d\rceil}+1)n)+poly(n)+2^{O(n^{k/d})}$.
\bigskip

It remains to prove Lemma \ref{l:ddistr}. The core change of the construction is the definition of circuits $C_h$, for $h\in {\cal F}_m$, which is given inductively by $$C_h:=\bigvee_{z\in\{0,1\}^{u}}\left(C_{h^z}\wedge \bigwedge_{i=1,\dots,u} \overbrace{\neg}^{1-z_i\text{ times}}x_{m-u+i}\right),$$
where $h^z:=h(x_1,\dots,x_{m-u},z_1,\dots,z_u)\in {\cal F}_{m-u}$, $C_{h(b)}=h(b)$ for $b\in \{0,1\}^m$ and $u:=min\{\lceil n/d\rceil,m\}$.

Consequently, $D_g:=(C_{f\oplus g}\wedge C_{\neg g})\vee (C_{\neg (f\oplus g)}\wedge C_g)$ 
 is a circuit of depth $2d+3$. Recall that ${\bm h}=\overline{D_{{\bm g}}}$.

The definition of $\bm \delta$ is modified analogously, following the structure of $D_g$. 
 Set $v:=\lceil n/d\rceil$ and $w:=n\ mod \ \lceil n/d\rceil$ if $n\ne 0\ mod\ \lceil n/d\rceil$, $w:=\lceil n/d\rceil$ otherwise. Let $m\in [d]\cup\{\oplus\}, t\in \{0,1\}^{r}\cup\{\vee\}$, with $r:=1$ if $m=\oplus$, $r:=v$ if $1<m\le d$, $r:=w$ if $m=1$, and ${\bm g_{w+(m-1)v}}\in{\cal F}_{w+(m-1)v}$, with ${\bm g_{w+(\oplus-1)v}}:={\bm g_n}\in {\cal F}_{w+(\oplus-1)v}:={\cal F}_n$, be chosen uniformly at random and independently up to the dependence which is described explicitly. Then,

\[{\bm \delta}:=
\begin{cases}
\delta_{\vee}(\overline{C_{f\oplus {\bm g_n}}\wedge C_{\neg {\bm g_n}}}, \overline{C_{\neg (f\oplus {\bm g_n})}\wedge C_{{\bm g_n}}}) & \text{if } m=\oplus; t=\vee;\\
\delta_{\wedge}(\overline{C_{t\oplus f\oplus {\bm g_n}}}, \overline{C_{t\oplus \neg {\bm g_n}}}) & \text{if } m=\oplus; t\in \{0,1\};\\
\delta_{\vee}(\overline{C_{{\bm g_{w+(m-1)v}}^{1\dots 1}}\wedge \bigwedge_{i=1,\dots,v} x_{w+(m-2)v+i}},\dots & \\
\quad\quad\dots, \overline{C_{{\bm g_{w+(m-1)v}}^{0\dots 0}}\wedge \bigwedge_{i=1,\dots,v} \neg x_{w+(m-2)v+i}}) & \text{if } 1<m\le d; t=\vee;\\
\delta_{\wedge}(\overline{C_{{\bm g_{w+(m-1)v}}^{t}}}, \overline{ \underbrace{\neg}_\text{$1-t_1$ times} x_{w+(m-2)v+1}},\dots & \\ 
\quad\quad\quad\quad\quad\quad\hspace{40pt} \dots, \overline{ \underbrace{\neg}_\text{$1-t_{v}$ times} x_{w+(m-1)v}}) & \text{if } 1<m\le d; t=\{0,1\}^{v};\\
\delta_{\vee}(\overline{C_{{\bm g_{w}}^{1\dots 1}}\wedge \bigwedge_{i=1,\dots,w} x_{i}},\dots & \\
\quad\quad\dots, \overline{C_{{\bm g_{w}}^{0\dots 0}}\wedge \bigwedge_{i=1,\dots,w} \neg x_{i}}) & \text{if } m=1; t=\vee;\\
\delta_{\wedge}(\overline{C_{{\bm g_{w}}^{t}}}, \overline{ \underbrace{\neg}_\text{$1-t_1$ times} x_{1}},\dots, \overline{ \underbrace{\neg}_\text{$1-t_{w}$ times} x_{w}}) & \text{if } m=1; t=\{0,1\}^{w}.
\end{cases}
\]

Now, for each $x$, there are $3+4(2^{v}+1)d$ positions of gates in $D_g$ such that for each $g\in {\cal F}_n$ satisfying $\overline{D_g}(x)\ne f(x)$ an error set corresponding to one of these positions covers $x$. This is because each $x$ is consistent with exactly one disjunct on the path from the output of $D_g$ to its inputs. The final analysis thus differs only in that instead of $3(n+1)$ we have $\le (2^{v}+1)(d+1)$ types. \qed 
\bigskip

Theorem \ref{t:damlim} does not rule out the possibility of obtaining \NC lower bounds via the approximation method.

\begin{problem}[\NC lower bounds via the approximation method]\label{pr:nc} Let $\cal M$ be a legitimate model as in Theorem \ref{t:damlim} and $f\in {\cal F}_n$. Is $\rho_d(f,{\cal M})\le 2^{n^{O(1/d)}}$, for each sufficiently big $d$? Does the upper bound hold for all $f$ computable by nondeterministic circuits of p-size?
\end{problem}

\noindent Note that a straightforward application of approximation models from the Razborov-Smolensky lower bound does not show that $\rho_d(f,{\cal M})\le 2^{n^{O(1/d)}}$ fails for some $f$: Since $n$-degree polynomials compute all functions $f\in {\cal F}_n$, to prove the existence of a hard function for $l^d$-degree polynomials (approximating depth $d$ circuits), we need $l<n^{1/d}$. As individual error sets in the model consisting of $l^d$-degree polynomials might cover up to $1/2^l$ of inputs, the existence of a function $f$ such that $\rho_d(f,{\cal M})>2^l$ does not follow directly.
\bigskip

Even though Theorem \ref{t:damlim} might not be optimal it can be used to derive a nontrivial localization of the approximatiom method for constant-depth circuits.

\begin{theorem}[Localizability of the approximation method for constant-depth circuits]\label{t:dloc} Let $\cal M$ be any legitimate model $\cal M$ of Boolean circuits over $\bigcup_{a\ge 2}\{\neg, \wedge_{a},\vee_{a}\}$ such that $\overline{\neg}:=\neg$, and $f\in {\cal F}_n$. Suppose that $\rho_{d(2d+6)}(f,{\cal M})\ge s$, for $d\ge 1$. 

Then, $f$ is not computable by depth $d$ circuits of size $s-poly(kdn(2^{m/d}+1))$ 
 using connectives $\neg,\wedge_{a},\vee_{a}$, for $a\ge 2$, and $k$ arbitrarily powerful, possibly different and noncommutative, oracles of arity $m$.\footnote{Theorem \ref{t:dloc} holds also for Razborov's asymmetric 
definition of $\rho(f,{\cal M})$, if we conclude that the resulting lower bound holds only for circuits with monotone oracles and negations at the bottom. Further, the requirement on the monotonicity of oracles and negations at the bottom can be avoided by using slice functions and adjusting the parameters appropriately. 
} The constants in the $poly(\cdot)$-notation are universal and, in particular, independent of $d$.
\end{theorem}

Theorem \ref{t:dloc} does not need any projectivity 
 assumption, but just like in Theorem \ref{t:damlim} its conclusion might not be optimal. Note that showing that $f$ is not computable by $poly(n2^m)$-size circuits of depth $d$ implies that $f$ is not computable by p-size circuits of depth $d/3$ 
 with oracles of arity $m$ because we can replace such oracles by DNFs of size $2^{O(m)}$. Theorem \ref{t:dloc} shows that even $poly(n2^{m/d})$-size lower bounds localize. Is it possible to localize lower bounds of size $poly(n2^{m^{O(1/d)}})$? A positive answer to Problem \ref{pr:nc} could resolve the question.
\medskip

\proof We proceed as in the proof of Item 1 in 
 Theorem \ref{t:genloc} with some modifications. 

We will inductively associate each $e^i$ with a function $f^i\in {\cal M}$ so that each $x$ satisfying $e^j(x)\ne f^j(x)$, for some $j\le i$, is covered by one of $ i+Ko_id(2^{Km/d}+1)n^K$ error sets $\delta\in\Delta$, for an absolute constant $K$. Additionally, we will ensure that for each $f^i$ there is a circuit $D_i$ of depth $d_i(2d+6)$ such that $f^i=\overline{D_i}$ and each $x$ such that $e^j(x)\ne f^j(x)$, for some $j\le i$, is covered by an error set whose input functions are approximators of circuits of depth $\le d_{j}(2d+6)-1$. Here, $d_i$ is the depth of the subcircuit corresponding to $e^i$. 
 In particular, we will have 
$\rho_{d_i(2d+6)}(e^i,{\cal M})\le i+Ko_id(2^{Km/d}+1)n^K$. 

The base case and the inductive step when the $(i+1)$st gate is not an oracle are straightforward. 

Suppose that the $(i+1)$st gate $\circ(e^{i_1},\dots,e^{i_m})$ computes an oracle of arity $m$. Following the proof of Lemma \ref{l:ddistr} (with $f^{i_j}$'s instead of $x_j$'s, with $\overline{f^{i_j}}:=f^{i_j}$ and $f=\circ$), we obtain a  random variable ${\bm\delta}$ on $\Delta$ and a random variable ${\bm h}$ on $\cal M$ such that for each $x$, $$\Pr[{\bm h}(x)\ne \circ(f^{i_1},\dots,f^{i_m})(x)]\le 4(d+1)(2^{\lceil m/d\rceil}+1)\Pr[x\in {\bm \delta}].$$ Moreover, for each $h$ in the range of ${\bm h}$ there is a circuit $D$ of depth $\le 2d+3+(d_{i+1}-1)(2d+6)$ such that $h=\overline{D}$ and for each $\delta_{\circ}(f_1,\dots,f_t)$ in the range of $\bm \delta$ there are circuits $D_{\ell}$ of depth at most $2d+2+(d_{i+1}-1)(2d+6)$ such that $f_{\ell}=\overline{D_{\ell}}$.

Next, we simulate the case analysis from the proof of Theorem $\ref{t:genloc}$.

In Case I we consider $x$'s such that $\Pr[{\bm h}(x)\ne \circ(f^{i_1},\dots,f^{i_m})(x)]<1/6$ and use approximate majority instead of {\sf MAJ}. Ajtai \cite{Afla} constructed (monotone) p-size circuits of depth 3 accepting each input which contains $>3/4$ 1s and rejecting each input which contains $<1/4$ 1s. This yields $d_{i+1}(2d+6)$ depth circuits and $K_1n^{K_1}$ corresponding error sets covering all $x$'s considered in Case I such that $f^{i+1}(x)\ne\circ(f^{i_1},\dots,f^{i_m})(x)$. Here, $K_1$ is an absolute constant.

In Case II we cover the remaining $x$'s by $K_2d(2^{\lceil m/d\rceil}+1)n$ error sets, for an absolute constant $K_2$.

Cases I and II together with the inductive hypothesis imply that all $x$'s such that $f^j(x)\ne e^j(x)$, for some $j\le i+1$,
 are covered by $Ko_{i+1}d(2^{\lceil m/d\rceil}+1)n^K+(i+1)$ error sets.  Moreover, for each gate $e^j$, $j\le (i+1)$, $f^j$ is computed by an approximator of a circuit of depth $d_{j}(2d+6)$ and each $x$ such that $e^j(x)\ne f^j(x)$, for some $j\le i+1$, is covered by an error set whose inputs are approximators of circuits of depth $\le d_j(2d+6)-1$. 
 \qed

\section{Completeness: Fusion}\label{s:fusion}

\subsection{General circuits}\label{s:gfusion}

Theorem \ref{t:amlim} does not rule out the possibility of deriving superquadratic lower bounds via the approximation method by introducing inessential variables: We might be able to show that $f\notin \Circuit[s]$ by proving $s\le \rho(f',{\cal M})$ for $f'(x_1,\dots,x_N):=f(x_1,\dots,x_n)\in {\cal F}_N$ and $N=s$.

Razborov \cite{Ram1} showed that this strategy can be adapted so that it is complete. 
We present Karchmer's \cite{Kfus} interpretation of Razborov's construction for which Wigderson \cite{Wfus} coined the term {\em fusion} method.
\bigskip

\noindent Let $f\in {\cal F}_n$, $V:=f^{-1}(1), U:=f^{-1}(0)$. We aim to prove $f\notin \Circuit[s]$ by showing that each $s$-size circuit $C$ computing $f$ yields a rejecting computation of $C$ on an input from $V$.

The rejecting computions will be `filtered' from more abstract computations as follows. Given $g\in {\cal F}_n$, define $[[g]]:=g^{-1}(1)\cap U$. The sets $[[g]]$ can be seen as generalizations of values 0 and 1 with $\emptyset$ standing for 0 and $U$ for 1. Note that $[[f]]=\emptyset$ and for all $g,h\in {\cal F}_n$, $$[[g]]\cap [[h]]=[[g\wedge h]]\quad \text{and}\quad [[g]]\cup [[h]]=[[g\vee h]].$$ We say that $F\subseteq {\cal P}(U):=\{U'\mid U'\subseteq U\}$ is a {\em semi-filter}\footnote{In order theory, a filter on ${\cal P}(U)$ would satisfy also that $F(A)=1\wedge F(B)=1$ implies $F(C)=1$ for some $C\subseteq A\cap B$.}, if
\begin{itemize}
\item[1.] (nontriviality) $F(\emptyset)=0$, $F(U)=1$;
\item[2.] (monotonicity) $A\subseteq B\quad\Rightarrow\quad F(A)\le F(B)$.
\end{itemize}
\noindent A semi-filter $F$ `preserves' a pair of sets $(A,B)$, $A,B\subseteq U$, if $$F(A)=1\wedge F(B)=1\Rightarrow F(A\cap B)=1.$$ Otherwise, we say that $(A,B)$ covers $F$. Suppose that all negation gates in a circuit $C$ computing $f$ are at the bottom layer and that $F$ is a semi-filter with a string $v(F)\in\{0,1\}^n$ such that $$v(F)_i=F([[x_i]])\ne F([[\neg x_i]])\text{, for }i\in [n].$$ If $F$ preserves pairs $([[g]],[[h]])$, for all gates $g,h$ of $C$, then $F$ implicitly defines a rejecting computation of $C$ on input $v(F)$.

Let $F_0$ be an arbitrary set of semi-filters $F\subseteq {\cal P}(U)$ with $v(F)\in V$. Define $\rho_{F_0}(f)$ as the minimal number of pairs $(A,B)$, $A,B\subseteq U$ which cover all semi-filters $F\in F_0$. It follows directly from the definition that $$\rho_{F_0}(f)\le 2\cdot Size_{{\cal C}}(f),$$ where ${\cal C}$ is the set of all Boolean circuits over $\{\neg,\wedge_2,\vee_2\}$. The factor of 2 results from pushing negations to the bottom.

The term `fusion' refers to the process of `fusing' rejecting computations of $C$ on $U$ into a rejecting computation of $C$ on an input from $V$.
\bigskip

\noindent {\bf Fusion as the approximation method.} The fusion method is an instantiation of the approximation method, if we relax the definition of a legitimate model. To see that, we use Razborov's definition of the distance $\rho(f',{\cal M})$ where $f'\oplus g\subseteq \bigcup_i \delta_{\circ_i}(f^i_1,\dots,f^i_{m_i})$ is replaced by $$f'\backslash g\subseteq\bigcup_i \delta^{+}_{\circ_i}(f^i_1,\dots,f^i_{m_i})\quad\text{and}\quad g\backslash f'\subseteq\bigcup_i \delta^{-}_{\circ_i}(f^i_1,\dots,f^i_{m_i}).$$ Assume that $|V|\ge 2$ and let $F_0$ be a set of semi-filters $F\subseteq {\cal P}(U)$ with $v(F)\in V$ such that there are $F_1,F_2\in F_0$ with $v(F_1)\ne v(F_2)$. We want to define a legitimate model ${\cal M}$ of order $N:=n+\lceil\log |F_0|\rceil$. $\cal M$ will consist of functions $\overline{g}\in {\cal F}_N$, for all $g\in {\cal F}_n$, given by
\[\overline{g}(x,y):=
\begin{cases} 
g(x) & \text{if }v(y)\ne x,\\
y([[g]]) & \text{if }v(y)=x;
\end{cases}
\]
where $x\in \{0,1\}^n$ and $y\in \{0,1\}^{N-n}$ is identified with the $y$th semi-filter from $F_0$. In particular, $\{0,1,x_i,\neg x_i\mid i\in [n]\}\subseteq {\cal M}$. We relax on the requirement that $x_i\in {\cal M}$, for $i>n$. The connectives of $\cal M$ are \[\overline{\circ}(\overline{g},\overline{h}):=\overline{g\circ h},\] for $\circ\in\{\vee_2,\wedge_2\}$. As $F_0$ contains $F_1,F_2$ with $v(F_1)\ne v(F_2)$, we have $\overline{g}=\overline{h}\Rightarrow g=h$ and $\overline{\circ}$ is well-defined. Note that $\cal M$ depends on $f$.

We want to show that $\rho(f',{\cal M})=\rho_{F_0}(f)$. The function $g\in {\cal M}$ from the definition of $\rho(f',{\cal M})$ has to be $\overline{f}$ because error sets $\delta_{\circ}$ do not contain any input $(x,y)$ with $v(y)\ne x$. Further, by definition, $\overline{f}(x,y)\ne f'(x,y)\Leftrightarrow v(y)=x$ and $v(y)=x\Rightarrow \overline{f}(x,y)=0$. Therefore, $\rho(f',{\cal M})$ is the minimal number of pairs $g,h$ such that pairs $([[g]],[[h]])$ cover all semi-filters in $F_0$, i.e. $\rho(f',{\cal M})=\rho_{F_0}(f)$.
\medskip

\noindent {\bf {\em Genesis of fusion.}} The fusion method arises from the approximation method by a sequence of natural choices. We want to approximate each $g\in {\cal F}_n$ by a function $\overline{g}\in {\cal F}_N$. Given the correspondence we look for a model with simple approximating connectives $\overline{\circ}(\overline{g},\overline{h})=\overline{g\circ h}$. For this to be well-defined we want $\overline{g}=\overline{h}\Rightarrow g=h$, which we guarantee by setting $$\overline{g}(x,y):=g(x)\quad \text{if }P(x,y)$$ for some predicate $P$. If $\neg P(x,y)$, we set $\overline{g}(x,y):=F_y(g)$
for a functional $F_y:2^{2^n}\mapsto \{0,1\}$. $F_y$ should guarantee that $\overline{x_i}=x_i$ and $\overline{\neg x_i}=\neg x_i$ so $\neg P(x,y)$ is ``$F_y(x_i)$'s define $x$". By imposing the monotonicity of $F_y$ we make $\delta^{+}_{\vee}$ empty. Similarly, setting $F_y(f):=0$ simplifies the situation as it implies that $\overline{f}\backslash f'=\emptyset$ and restricts our attention to $U$.

\begin{theorem}[Completeness of the fusion method \cite{Ram1}]\label{t:fcomp} Let $f\in {\cal F}_n$ and assume that $f\notin\Circuit[Ks^3]$, for $s\ge n$. Then $\rho_{F_{\forall}}(f)\ge s$, where $F_{\forall}$ is the set of all semi-filters $F\subseteq {\cal P}(U)$ with $v(F)\in V$. Moreover, there is $F_0\subseteq F_{\forall}$ of size $|F_0|=2^{O(s|U|)}$ such that $\rho_{F_0}(f)\ge s$. Here, $K$ is an absolute constant.
\end{theorem}

\proof Assuming $\rho_{F_{\forall}}(f)\le s$, we want to construct an $O(s^3)$-size circuit computing $f$. We have $s$ pairs $(A_i,B_i)$, $A_i,B_i\subseteq U, i\in [s]$ covering all semi-filters in $F_{\forall}$. Let $F_z$, for $z\in \{0,1\}^n$, be the minimal subset of ${\cal P}(U)$ closed on the supersets, preserving all pairs $(A_i,B_i)$, for $i\in [s]$, pairs $([[x_i]],[[\neg x_i]])$, for $i\in [n]$, and such that for all $i\in [n]$, $$[[\underbrace{\neg}_{1-z_i\text{ times}} x_i]]\in F_z.$$

We claim that $$f(z)=1\quad\Leftrightarrow\quad \emptyset\in F_z.$$ If $f(z)=0$, then $Z:=\{U'\mid z\in U'\subseteq U\}$ witnesses that $\emptyset\not\in F_z$. If $f(z)=1$, then $\emptyset\in F_z$ as otherwise $v(F_z)\in V$ and $F_z$ would be a semi-filter preserving all pairs $(A_i,B_i), i\in [s]$.

It remains to construct an $O(s^3)$-size circuit deciding $\emptyset\in^{?}F_z$. 

Let $$A:=\bigcup_{i\in [s]}\{A_i,B_i,A_i\cap B_i\}\cup\bigcup_{i\in [n]}\{[[x_i]],[[\neg x_i]]\}\cup\{\emptyset\}.$$ For each $B\in A$ and $k\ge 0$, define a function $w^k_B$ inductively by
\[w^0_{B}:=
\begin{cases}
1 &\text{if for some }i\in [n], B=[[x_i]]\wedge z_i=1\text{ or }B=[[\neg x_i]]\wedge z_i=0,\\
0 & \text{otherwise}
\end{cases}
\]
$$w^{k+1}_B:=\bigvee_{C\subseteq B,C\in A} w^k_C\vee\bigvee_{j\in \{j\mid A_j\cap B_j=B\}} (w^k_{A_j}\wedge w^k_{B_j}) \vee\bigvee_{i\in [n]}(w^k_{[[x_i]]}\wedge w^k_{[[\neg x_i]]}).$$ Then, $\emptyset\in F_z\Leftrightarrow w^{|A|}_{\emptyset}=1$ and the definition of $w^{|A|}_{\emptyset}$ yields a circuit of size $O(s^3)$ computing $f$.
\medskip

The `moreover' part follows by noting that it suffices to consider $2^n$ semi-filters $F\in F_{\forall}$ for each possible set of $s$ pairs $(A_i,B_i), i\in [s]$.
  \qed

\bigskip
\def\pro{

\begin{proposition}[Je\v{r}\'{a}bek \cite{Japx}](in \APC)\label{really} 

\noindent 1.\ \ Let $X,Y\subseteq 2^n$ be definable by circuits, $s,t,u\leq 2^n$, $\epsilon, \delta,\theta, \gamma < 1, \gamma^{-1}\in Log $. Then

i)\ \ $X\preceq_{\gamma} Y$ or $Y\preceq_{\gamma} X$,

ii)\ \ $s\preceq_\epsilon X\preceq_\delta t\Rightarrow s<t+(\epsilon+\delta+\gamma)2^n$,

iii)\ \ $X\preceq_{\epsilon} Y\Rightarrow 2^n\backslash Y\preceq_{\epsilon+\gamma} 2^n\backslash X $,

iv)\ \ $X\approx_{\epsilon} s\wedge Y\approx_{\delta} t\wedge X\cap Y\approx_{\theta} u\Rightarrow X\cup Y\approx_{\epsilon+\delta+\theta+\gamma} s+t-u$.

\medskip

\noindent 2. (Disjoint union) Let $X_i\subseteq 2^n$, $i<m$ be defined by a sequence of circuits and $\epsilon,\delta\leq 1$, $\delta^{-1}\in Log$. If $X_i\preceq_\epsilon s_i$ for every $i<m$, then $\bigcup_{i<m} (X_i\times \{i\})\preceq_{\epsilon+\delta} \sum_{i<m} s_i$.
\end{proposition}

}

\noindent {\bf Feasibility of fusion.} In the approximation method, error sets $\delta_{\circ}$ consist of inputs of the target function. In the fusion method, the error sets $\delta_{\circ}$ consist of semi-filters. By Theorem \ref{t:amlim}, the number of semi-filters cannot be reduced below $2^{\Omega(s)}$, if the fusion method is to prove an $s$-size lower bound.

Razborov \cite{Ram2} came up with an elegant generalization of the approximation method where the error sets are subsets of a set $S$ which is unrelated to the inputs of the target function and has size just $|S|=O(s^3\log^2 s)$. A drawback of his construction is that it is nonconstructive and not known to be complete - it proves $s$-size lower bounds for {\em almost all} functions, for suitable $s$. 

The motivation for the construction from \cite{Ram2} was the question of finding a version of the approximation method capable of proving strong lower bounds which would be suitable for formalizations in bounded arithmetic \SB. Informally, \SB is a theory of polynomial-time reasoning and does not allow to operate with objects which cannot be described by bit-strings of polynomial-size, cf. \cite{Bba}. The polynomial-size is considered w.r.t. an initial parameter which is typically $s(n)$, where $n$ is the number of inputs of the target function and $s(n)$ is the size of the lower bound we aim to prove, but it is possible to consider also, say, $2^n$ as the initial parameter, if the lower bound is formulated w.r.t. the truth-table of the target function. The latter formulation corresponds to the setting in the natural proofs and we refer to it as the `truth-table' formalization, see \cite{MP}.

If we want to formalize an $s$-size lower bound for $f\in {\cal F}_n$ in \SB using the truth-table formalization, it suffices to show that for each set of $s$ pairs $(A,B),A,B\subseteq U$, there is a suitable semi-filter preserving the set. In the truth-table formalization, \SB can operate only with objects described by $2^{O(n)}$ bits. Nevertheless, semi-filters from $F_0$ in Theorem \ref{t:fcomp} can be chosen so that they are predicates computable by circuits of size $2^{O(n)}$. That is, the size of notions appearing in the fusion method does not present an obstacle for proving circuit lower bounds in \SB using the truth-table formalization. 

With the more succinct formalization of lower bounds in \SB, where the initial parameter is $s(n)$ instead of $2^n$, each semi-filter needs to be described by $poly(s)$ bits. This leads to the following adaptation of the fusion method, which we refer to as the {\em antichecker fusion}.

Lipton and Young \cite[Theorem 6]{LY} proved that for each sufficiently big $n$ and each function $f\in {\cal F}_n$ such that $f\notin \Circuit[s^7]$, $s\ge n$, there is a set $S\subseteq \{0,1\}^n$ of size $|S|=poly(s)$ such that no $s^3$-size circuit computes $f$ on $S$. The set $S$ is the set of {\em anticheckers} of $f$. Consider the fusion method modified so that instead of sets $U,V$ we use $U\cap S, V\cap S$. If we denote by $\rho^S_{F_0}$ the resulting version of $\rho_{F_0}$, we have $\rho^S_{F_0}(f_n)\le 2\cdot Size_{\cal C}(f)$, for each set $F_0$ of semi-filters $F\subseteq {\cal P}(U\cap S)$ with $v(F)\in V\cap S$. Moreover, by the proof of Theorem \ref{t:fcomp}, for some $F_0$ with $|F_0|=2^{poly(s)}$, $f\notin \Circuit[Ks^7]$ implies $\rho^S_{F_0}(f)\ge s$. This proves the following. 

\begin{theorem}[Completeness of the antichecker fusion]\label{t:afcomp} Let $f\in {\cal F}_n$, for a sufficiently big $n$, and assume that $f\notin\Circuit[Ks^7]$, for $K^{1/7}s\ge n$. Then there is a set $F_0$ of semi-filters $F\subseteq {\cal P}(U\cap S)$ with $v(F)\in V\cap S$ and a set $S$ of anticheckers of $f$ such that $|F_0|=2^{poly(s)}$ and $\rho^S_{F_0}(f)\ge s$. Here, $K$ is an absolute constant.
\end{theorem}

\noindent The approximation model $\cal M$ corresponding to the antichecker fusion from Theorem \ref{t:afcomp} uses $N=poly(s)$ and semi-filters from $F_0$ are predicates computable by $poly(s)$-size circuits. We can consider a more efficient approximation model ${\cal M}_S$ consisting of functions $\overline{g}(x,y)\in {\cal M}$ restricted to $S$ on $x$'s. That is, ${\cal M}_S$ consists of partial Boolean functions - the approximation method straightforwardly generalizes to this setting. Functions $\overline{g}\in {\cal M}_S$ are computable by $poly(s)$-size circuits. Further, for each $f_1,f_2\in {\cal M}_S$, the predicate $(x,y)\in^{?}\delta^{+}_{\wedge}(f_1,f_2)$ is computable by a $poly(s)$-size circuit. Therefore, if we ignore the question of generating anticheckers efficiently and consider only the size of the notions involved, \SB is perfectly capable of proving circuit lower bounds by estimating $\rho(f',{\cal M}_S)$.

A nonconstructive element of the model ${\cal M}_S$ is that we do not have a p-time algorithm which would output the set of anticheckers given a function $f\in {\cal F}_n$. For $f=\SAT$, the set of anticheckers w.r.t. $s=poly(n)$ can be generated in p-time given $1^n$ under the assumption of the existence of a one-way function secure against nonuniform p-size circuits and a function in {\sf E} hard for subexponential-size circuits \cite{MP}. 

The complication with generating anticheckers disappears if we interpret the antichecker fusion as a {\em fusion method for partial Boolean functions}: Given $f:S\mapsto \{0,1\}$, $S\subseteq \{0,1\}^n$, which is not computable by any $Ks^7$-size circuit, we have $s\le \rho^S_{F_0}(f)$ and $\rho^S_{F_0}(f)/2$ is a lower bound on the size of the smallest circuit computing $f$. 

\subsection{Constant-depth circuits} 

We now adapt the fusion method to the case of constant-depth circuit lower bounds. To make this work we break the monotonicity of semi-filters.
\bigskip

Let $f\in {\cal F}_n, V:=f^{-1}(1), U:=f^{-1}(0)$. We say that $F=(F^0,\dots,F^d)$, where $F^0\subseteq\dots\subseteq F^d\subseteq {\cal P}(U)$ is a $d$-semifilter with $v(F)\in V$, if $F^d(\emptyset)=0$, $F^0(U)=1$ and
\begin{itemize}
\item[1.] (initial sets) $v(F)_i=F^0([[x_i]])\ne F^0([[\neg x_i]])$, for $i\in [n]$;
\item[2.] ($d$-monotonicity) For $1\le k\le d$, $F^{k-1}(B)=1\wedge B\subseteq A\Rightarrow F^{k}(A)=1.$
\end{itemize}

A $d$-semifilter $F$ $k$-preserves a tuple $(A_1,\dots, A_t), A_i\subseteq U$, for $1\le k\le d$, if $$\bigwedge_{i=1,\dots,t} F^{k-1}(A_i)=1\Rightarrow F^k(\bigcap_{i=1,\dots, t} A_i)=1.$$ Otherwise, we say that $(A_1,\dots, A_t)$ $k$-covers $F$. 

Suppose that $C$ is a $(d+1)$-depth circuit over $\bigcup_{a=2,\dots,t} \{\neg,\wedge_a,\vee_a\}$ with negation gates at the bottom layer. Further, assume that $C$ computes $f$ and that $F$ is a $d$-semifilter with $v(F)\in V$. If for each $1\le k\le d$, $F$ $k$-preserves tuples $([[g_1]],\dots,[[g_t]])$, for all gates $g_1,\dots,g_t$ of $C$ computable by $k$-depth subcircuits of $C$, then $F$ implicitly defines a rejecting computation of $C$ on input $v(F)$. 

Let $F_0$ be a set of $(d-1)$-semifilters $F$ with $v(F)\in V$, for $d\ge 2$. Define $\rho_{F_0,d,t}(f)$ as the minimal number of tuples $(A_1,\dots, A_{t}), A_i\subseteq U,$ such that for each $F\in F_0$, there is $1\le k<d$ such that one of the tuples $k$-covers $F$. If such tuples do not exist, we set $\rho_{F_0,d,t}(f):=\infty$. Then, $\rho_{F_0,d,t}(f)/2$ is a lower bound on the size of a smallest $d$-depth circuit over $\bigcup_{a=2,\dots,t} \{\neg,\wedge_a,\vee_a\}$ computing $f$.

\begin{theorem}[Completeness of fusion for constant-depth circuit lower bounds]\label{t:dfcomp} Let $f\in {\cal F}_n$ and assume that no $(2d+1)$-depth $Kd(st+n)^2$-size circuit over $\bigcup_{a=2,\dots,s(t+2)+2n+2} \{\neg,\wedge_a,\vee_a\}$, for $d,n,t\ge 2$, computes $f$. Then, $\rho_{F_{\forall},d,t}(f)\ge s$, where $F_{\forall}$ is the set of all $(d-1)$-semifilters $F$ with $v(F)\in V$. Here, $K$ is an absolute constant.
\end{theorem}

\proof We proceed as in the proof of Theorem \ref{t:fcomp}. Assuming $\rho_{F_{\forall},d,t}(f)\le s$, we want to construct an $O(d(st+n)^2)$-size $(2d+1)$-depth circuit computing $f$. We have $s$ tuples $(A^i_1,\dots, A^i_{t})$, $i\in [s]$, covering all $F\in F_{\forall}$. 

Define $F_z=(F^0_z,\dots,F^{d-1}_z)$, where $F^0_z\subseteq\dots\subseteq F^{d-1}_z \subseteq{\cal P}(U)$, for $z\in \{0,1\}^n$ as follows. $F_z^0$ consists of the set $U$ and sets $[[\underbrace{\neg}_{1-z_i\text{ times}} x_i]]$, for all $i\in [n]$. For $k\ge 1$, define $F^{k-1}_z\subseteq F_z^k\subseteq {\cal P}(U)$ by extending $F^{k-1}_z$ by all supersets of sets in $F^{k-1}_z$ and by `$k$-preserving' tuples $(A^i_1,\dots, A^i_{t})$, for $i\in [s]$, consisting of sets from $F^{k-1}_z$.

We claim that $f(z)=0\Leftrightarrow \emptyset\notin F_z^{d-1}$. If $f(z)=0$, then $z$ is included in all sets in $F_z^{d-1}$ so $\emptyset\notin F_z^{d-1}$. 
If $f(z)=1$, then $\emptyset\in F_z^{d-1}$ as otherwise $v(F_z)\in V$ and $F_z$ would be a $(d-1)$-semifilter $k$-preserving all tuples $(A^i_1,\dots, A^i_{t}), i\in [s]$, for all $1\le k<d$.

It remains to construct a circuit deciding $\emptyset\in^{?}F_z^{d-1}$. 
Let $$A:=\bigcup_{i\in [s]}\{A^i_1,\dots,A^i_{t},\bigcap_{j=1,\dots,t} A^i_j\}\cup \bigcup_{i\in [n]}\{[[x_i]],[[\neg x_i]]\}\cup\{\emptyset,U\}.$$ For each $B\in A$ and $k\ge 0$, define a function $w^k_B$ inductively by
\[w^0_{B}:=
\begin{cases}
1 &\text{if }\exists i\in [n], B=[[x_i]]\wedge z_i=1\text{ or }B=[[\neg x_i]]\wedge z_i=0\text{ or }B=U,\\
0 & \text{otherwise}
\end{cases}
\]
$$w^{k+1}_B:=\bigvee_{C\subseteq B,C\in A} w^k_C\vee\bigvee_{i\in \{i\mid B=\bigcap_{j=1,\dots,t} A^i_j\}} \left(\bigwedge_{j=1,\dots,t} w^k_{A^i_j}\right).$$
By definition, $B\in F^k_z\Leftrightarrow w^k_B=1$, for $k\in\{0,1\}$ and $B\in A$. For each $C\in F^{k}_z$, $1\le k<d$, with $C\subseteq B\in A$, $C$ was added to $F^k_z$ either as a superset of some $D\in F^{k-1}_z$, in which case $B\in F^k_z\cap A$, or $C$ was added to $F^k_z$ in order to preserve a tuple, in which case $C\in A$. This allows us to prove by induction that the previous equivalence holds for $0\le k<d$. In particular, $\emptyset\in F_z^{d-1}\Leftrightarrow w^{d-1}_{\emptyset}=1$. Finally, by the construction of $w^{d-1}_B$, there are $(2d+1)$-depth $O(d(st+n)^2)$-size circuits over $\bigcup_{a=2,\dots,s(t+2)+2(n+1)}\{\neg,\vee_a,\wedge_a\}$ computing $w^{d-1}_B$. 
  \qed
\bigskip

\def\dapproxm{
\noindent {\bf \textcolor{red}{Incorrect} Approximation models for constant-depth fusion.} To construct approximation models capturing the adaptation of fusion to constant-depth circuits, we make a couple of adjustments. First, we redefine $\rho_{F_0,d,t}$ so that the tuples $(A_1,\dots,A_{t})$ consist of sets $A_i\subseteq U$ such that $A_i=[[g]]$ for a function $g$ computable by a $(d-1)$-depth circuit over $\bigcup_{a=2,\dots,t}\{\neg,\wedge_a,\vee_a\}$. Additionally, we will consider only sets $F_0$ consisting of $(d-1)$-semifilters $F$ which satisfy $F^k([[x_i]])\ne F^k([[\neg x_i]])$, for all $0\le k<d$. The resulting measure, denoted $\rho_{F_0,d,t}'$, is still a lower bound for $d$-depth circuits and it satisfies completeness from Theorem \ref{t:dfcomp} with a minor change of parameters: In the proof of Theorem, \ref{t:dfcomp} we additionally $k$-preserve pairs $([[x_i]],[[\neg x_i]])$, $i\in [n]$, in each $F_z^k$, $1\le k<d$, and include one more step in which we add $\emptyset$ to $F_z^{d-1}$, if $F^{d-1}([[x_i]])=F^{d-1}([[\neg x_i]])$, for some $i\in [n]$.

Second, we consider a multidimensional approximation method in which legitimate models are subsets of ${\cal F}^m_n$, the set of all $n$-input Boolean functions with $m$ output bits. Legitimate models of the multidimensional approximation method are required to include $I:=\{(0)^m,(1)^m,(x_i)^m,(\neg x_i)^m\mid i\in [n]\}$, where $(g)^m:=(g,\dots,g)\in {\cal F}^m_n$, for $g\in {\cal F}_n$. The definition of approximating circuits remains the same except that we use functions from $I$ in the initial step. For errors sets, we have $$\delta^{+}_{\circ}(f_1,\dots,f_k):=\bigcup_{i=1,\dots,m}(\circ(f_1|_i,\dots,f_k|_i)\backslash\overline{\circ}(f_1|_i,\dots,f_k|_i)),$$ where $f_j|_i\in {\cal F}_n$ is the restriction of $f_j\in {\cal F}^m_n$ to the $i$th output bit. Similarly for $\delta^{-}_{\circ}$. The distance $\rho^m(f,{\cal M})$, for $f\in {\cal F}_n$, is defined as a direct analogue of Razborov's version of $\rho$ with $\bigcup_{i=1,\dots,m}(g|_i\backslash f)$ instead of $g\backslash f$. Proposition \ref{p:approxbound} can be adapted to $\rho^m$. 
Following the definition of $\rho_d$, the distance $\rho^m_d$ is defined analogously by requiring that for the function $g$ in the definition of $\rho^m$ there is a $d$-depth ${\cal C}$-circuit $B$ with $\overline{B}=g$ and the input circuits of error sets have depth $\le d-1$.

We are now ready to show that the multidimensional approximation method captures $\rho_{F_0,d,t}'$.

Assume that $|V|\ge 2$ and let $F_0$ be a set of $(d-1)$-semifilters $F$ with $v(F)\in V$, $d\ge 2$, such that there are $F_1,F_2\in F_0$ with $v(F_1)\ne v(F_2)$. Define a model ${\cal M}$ of order $N:=n+\lceil\log |F_0|\rceil$ as follows. $\cal M$ consists of functions $\overline{g}\in {\cal F}_N^{d-1}$, for all $g\in {\cal F}_n$, 
given by
\[\overline{g}(x,y):=
\begin{cases} 
(g(x),\dots,g(x)) & \text{if }v(y)\ne x,\\
(y^1([[g]]),\dots,y^{d-1}([[g]])) & \text{if }v(y)=x;
\end{cases}
\]
where $x\in \{0,1\}^n$ and $y^k\in \{0,1\}^{N-n}$ is identified with the set $F^k$ of the $y$th semi-filter $F=(F^0,\dots,F^{d-1})\in F_0$. 
In particular, $\{(0)^{d-1},(1)^{d-1},(x_i)^{d-1},(\neg x_i)^{d-1}\mid i\in [n]\}\subseteq {\cal M}$. Here, we use $F^k([[x_i]])\ne F^k([[\neg x_i]])$, for all $1\le k<d$. The connectives of $\cal M$ are \[\overline{\circ}(\overline{g_1},\dots,\overline{g_a}):=\overline{\circ(g_1,\dots,g_a)},\] for $\circ\in\{\vee_a,\wedge_a\},a=2,\dots, t$. As $F_0$ contains $F_1,F_2$ with $v(F_1)\ne v(F_2)$, we have $\overline{g}=\overline{h}\Rightarrow g=h$ and $\overline{\circ}$ is well-defined.

We want to show that $\rho^{d-1}_d(f',{\cal M})=\rho_{F_0,d,t}(f)$, for $f'(x,y):=f(x)$. The function $g\in {\cal M}$ from the definition of $\rho^{d-1}_d(f',{\cal M})$ has to be $\overline{f}$ because error sets $\delta_{\circ}$ do not contain any input $(x,y)$ with $v(y)\ne x$. Further, by definition, $\overline{f}(x,y)\ne (f'(x,y))^{d-1}\Leftrightarrow v(y)=x$ and $v(y)=x\Rightarrow \overline{f}(x,y)=(0)^{d-1}$. Therefore, $\rho^{d-1}_d(f',{\cal M})$ is the minimal number of tuples $([[g_1]],\dots,[[g_{t}]])$, where each $g_i$ is computable by a $(d-1)$-depth circuit over.., such that each $F\in F_0$ is $k$-covered by one of the tuples, for some $1\le k<d$. That is, $\rho^{d-1}_d(f',{\cal M})=\rho_{F_0,d,t}(f)$.

We want $(d-1)$-semifilters $F_0$ to satisfy an additional property: $F^{k-1}([[g]])=0$, for $g$ computable by a $(k-1)$-depth circuit over.. implies $F^{k}([[g]])=0$, for $1\le k<d$. The property is satisfied by $F_0$ w.l.o.g. This is because if we have $F_0$ such that for each set of $s$ tuples some $F\in F_0$ preserves the set, we can obtain $F'$ preserving the set and satisfying the property: if g was assigned 1 at some point it had to represent a computation of a circuit, but this are w.l.o.g. minimalized. (unclear)
\bigskip}

\noindent {\bf Random restrictions as the approximation method.} 
By Theorem \ref{t:dfcomp}, essentially every lower bound for constant-depth circuits can be turned into a lower bound obtained by the fusion method for constant-depth circuits. In particular, this holds for lower bounds based on the method of random restrictions.


The method of random restrictions proceeds by showing that each small circuit can be trivialized by a partial restriction, which implicitly defines an incorrect computation of the circuit w.r.t. a suitable target function. This is similar to the construction of semi-filters in the fusion method. Moreover, the number of partial restrictions needed to trivialize all $poly(s)$-size constant-depth circuits is just $poly(s)$. This suggests that it might be possible to formalize random restrictions in the framework of the approximation method with just $O(\log s)$ inessential variables. Such formalization would allow us to conclude that random restrictions are subject to the localizability from Theorem \ref{t:dloc} (with the asymmetric definition of $\rho_d$). However, at this point we have not even constructed legitimate approximation models corresponding to the fusion method for constant-depth circuits.




It would be interesting to formulate random restrictions also in terms of approximation models with $poly(s)$ inessential variables with an addition property similar to 0-projectivity. Such models could be localizable similarly as in Theorem \ref{t:dloc} as well.
\def\nonstand{

\section{Non-standard models}

A non-standard model of arithmetic defines similar circuits (the circuit satisfied the same properties but is not isomorphic to standard circuit in $\mathbb{N}$). We will show how to use these similar circuit to construct an approximation model. Then we compare approximation models based on non-standard models to Matte/Ajtai's approach to proving complexity lower bounds via a construction of non-standard models. }

\section*{Acknowledgement}

I thank Lijie Chen, Rahul Santhanam, Bruno Cavalar, Shuichi Hirahara and Jan Kraj\'i\v{c}ek 
 for feedback on a draft of the paper, and Susanna de Rezende with Erfan Khaniki for helpful discussions on the approximation method. 
I thank Navid Talebanfard for bringing \cite{BS} to my attention.
J\'an Pich received support from the Royal Society University Research Fellowship URF$\backslash$R1$\backslash$211106.
This project has received funding from the European Union's Horizon 2020 research and innovation programme under the Marie Sk\l{}odovska-Curie grant agreement No 890220. For the purpose of Open Access, the author has applied a CC BY public copyright licence to any Author Accepted Manuscript version arising from this submission. 
\medskip

\noindent\fbox{\includegraphics[width=48pt,height=31pt]{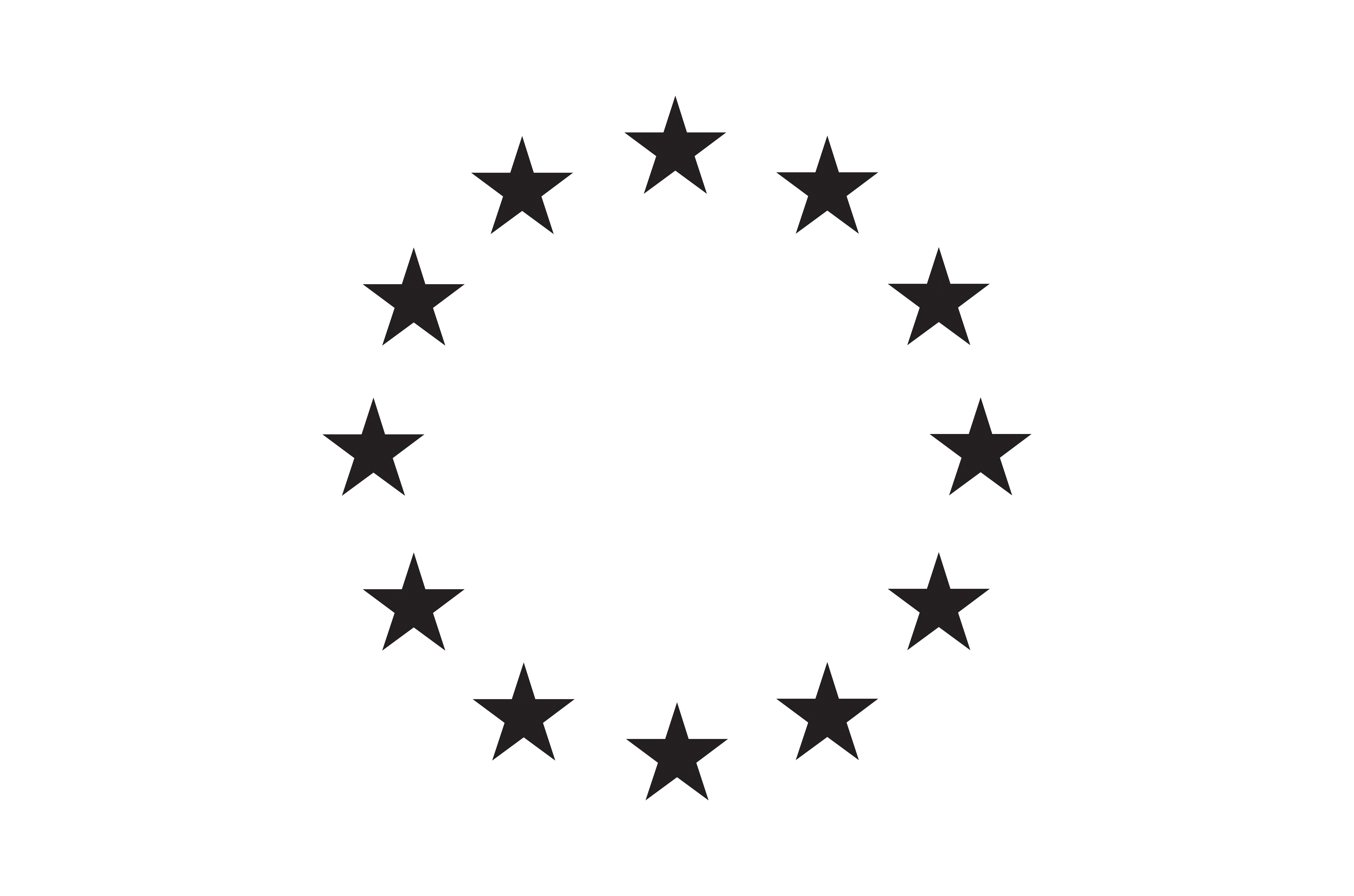}}

\end{document}